\documentclass{aa}
\usepackage[dvips]{graphicx}
\newcommand{\gtrsim}{\mathrel{\hbox{\rlap{\hbox{\lower4pt\hbox{$\sim$}}}\hbox{$>$}}}}
\newcommand{\lesssim}{\mathrel{\hbox{\rlap{\hbox{\lower4pt\hbox{$\sim$}}}\hbox{$<$}}}}

\def\MCN{\mbox{CH$_3$CN}}

\def\MCNII{CH${_3}^{13}$CN}

\def\HII{H{\sc ii} }

\def\UC{UC~H{\sc ii}}

\def\kms{\mbox{km~s$^{-1}$}}

\def\Vlsr{$V_{\rm LSR}$}

\def\mjy{~mJy~beam$^{-1}$}

\def\jdo{\mbox{12--11}}

\def\jdu{\mbox{2--1}}
\def\juz{\mbox{1--0}}
\def\jtd{\mbox{3--2}}
\def\jsc{\mbox{6--5}}

\begin{document}
\title{Rotating toroids in G10.62$-$0.38, G19.61$-$0.23, and G29.96$-$0.02\thanks{Based on observations carried out with the IRAM
Plateau de Bure Interferometer.  IRAM is supported by INSU/CNRS
(France), MPG (Germany) and IGN (Spain).}} 
\author{M.\ T. Beltr\'an\inst{1} \and R.\ Cesaroni \inst{1} \and R.\ Neri \inst{2}
\and C.\ Codella \inst{1}}
\institute{
INAF, Osservatorio Astrofisico di Arcetri, Largo E.\ Fermi 5,
50125 Firenze, Italy
\and
IRAM, 300 Rue de la Piscine, F-38406 Saint Martin d'H\`eres, France} 


\offprints{M.\ T.\ Beltr\'an, \email{mbeltran@arcetri.astro.it}}
\date{Received date; accepted date}

\titlerunning{Rotating toroids in HMCs}
\authorrunning{Beltr\'an et al.}

\abstract
{In recent years, we have detected clear evidence of rotation in more than 5 hot molecular cores (HMCs).
Their identification is confirmed by the fact that the rotation axes are parallel to the axes of the
associated bipolar
outflows. We have now pursued our investigation by extending the sample to 3
known massive cores, G10.62$-$0.38,  G19.61$-$0.23, and G29.96$-$0.02.}
{We wish to make a thorough study of the structure and kinematics of HMCs and corresponding molecular outflows to reveal possible velocity gradients
indicative of rotation of the cores.}
{We carried out PdBI observations at 2.7 and 1.4~mm of gas and dust with angular resolutions of
$\sim$2$''$--3$''$, and $\sim$1$''$--2$''$, respectively. To trace both rotation and expansion, we
simultaneously observed CH$_3$CN, a typical HMC tracer, 
and $^{13}$CO, a typical outflow tracer.}
{The \MCN~(12--11) observations have revealed the existence of clear velocity gradients in the three
HMCs oriented perpendicular to the direction of the bipolar outflows. For G19 and G29 the molecular outflows have
been mapped in $^{13}$CO. The gradients have been interpreted as rotating toroids. The rotation
temperatures, used to derive the mass of the cores, have been obtained by means of the rotational 
diagram method, and lie in the range of 87--244~K.
The diameters and masses of the toroids lie in the range of 4550--12600~AU, and 28--415~$M_\odot$, respectively. 
Given that the dynamical masses are 2 to 30 times smaller than the masses of the cores (if the
inclination of the toroids with respect to the plane of the sky is not much smaller than $45\degr$),
 we suggest that the toroids could be accreting onto the embedded cluster. For G19 and G29, the collapse is also suggested by the redshifted
absorption seen in the $^{13}$CO~(2--1) line. We infer that infall onto the embedded (proto)stars must proceed
with rates of $\sim$10$^{-2}$~$M_\odot$ yr$^{-1}$, and on timescales of the order of
$\sim$4$\times$10$^3$--10$^4$ yr. The infall rates derived for G19 and G29 are two orders of magnitude greater than the accretion rates
indirectly estimated from the mass loss rate of the corresponding outflows. This suggests that the material in
the toroids is not infalling onto a single massive star, responsible for the corresponding molecular
outflow, but onto a cluster of stars.}
{}

\maketitle

\section{Introduction}

How do massive ($>$ 8~M$_\odot$) stars form? Do they form like lower-mass stars or in a
fundamentally different way? What is their connection to star cluster formation? These are crucial
questions with profound implications for many areas of astrophysics including high-redshift
Population III star formation, galaxy formation and evolution, galactic center environments and
super-massive black hole formation, star and star cluster formation, and planet formation around
stars in clusters. 

\begin{table*}
\caption[] {Phase centers, distances, luminosities, and LSR velocities of the sources}
\label{table_pos}
\begin{tabular}{lccccccc}
\hline
&&\multicolumn{2}{c}{Position$^{a}$} & \\
 \cline{3-4} 
&&\multicolumn{1}{c}{$\alpha({\rm J2000})$} &
\multicolumn{1}{c}{$\delta({\rm J2000})$} &
\multicolumn{1}{c}{$L_{\rm bol}^b$} &
\multicolumn{1}{c}{$V_{\rm LSR}$} &
\multicolumn{1}{c}{$d$} \\
\multicolumn{1}{c}{Core} &
\multicolumn{1}{c}{IRAS PSC name} &
\multicolumn{1}{c}{h m s}&
\multicolumn{1}{c}{$\degr$ $\arcmin$ $\arcsec$} &
\multicolumn{1}{c}{($L_\odot$)} &
\multicolumn{1}{c}{(\kms)} &
\multicolumn{1}{c}{(kpc)} & 
\multicolumn{1}{c}{Refs.$^{c}$}\\
\hline
G10.62$-$0.38 &18075$-$1956 &18 10 28.650 &$-19$ 55 49.50 &0.4$\times$10$^6$ &$-$2.0$\phantom{.}$ &3.4 &1\\
G19.61$-$0.23 &18248$-$1158 &18 27 38.145 &$-11$ 56 38.49 &2.2$\times$10$^6$ &41.6 &12.6 &2\\ 
G29.96$-$0.02 &18434$-$0242 &18 46 03.955 &$-02$ 39 21.87 &3.2$\times$10$^5$ &98.9 &3.5 &3\\
\hline

\end{tabular}

 $^a$ Coordinates of the phase center of the observations. \\
 $^b$ The bolometric luminosities were calculated by integrating the 
 IRAS flux densities. The contribution from longer wavelengths was taken into account by extrapolating according
to a black-body function that peaks at 100~$\mu$m and has the same
flux density as the source at this wavelength. \\
 $^c$ References for the distances: 1: Blum et al.~(\cite{blum01}); 2: Kolpak et 
 al.~(\cite{kolpak03}); 3: L.\ Moscadelli (2010, private communication)    \\

\end{table*}

In contrast to our fairly mature understanding of isolated low-mass star
formation, grasping the details of high-mass star formation has been extremely
difficult. High-mass stars are rare objects that form very quickly. What is
more, the closest  high-mass star forming regions are located further away than
the nearest low-mass star forming regions. This, together with the fact that
massive stars form in crowded and very obscured regions, implies that the study
of the massive star formation process encounters some observational limitations
that sometimes cannot be circumvented. Furthermore, the interplay between
gravity, turbulence, radiation pressure, ionization and hydromagnetic outflows
is also expected to be more complicated in massive star forming region than in
low-mass ones.  From a theoretical point of view, three-dimensional radiation
hydrodynamic simulations by Krumholz et al.~(\cite{krumholz09}) have shown that
stars as massive as $\sim$40~$M_\odot$ may form through disk accretion and that
radiation pressure is not a barrier to form even more massive stars.  However,
these simulations have also shown that the instabilities that allow accretion to
continue lead to fragmentation and the formation of small multiple systems,
which could prevent the formation of stars with masses well beyond
$\sim$40~$M_\odot$. Therefore, from a theoretical point of view, the formation
process of the most massive stars still remains unknown. However, in a very
recent work, Kuiper et al.~(\cite{kuiper10}) form stars in excess of 100~$M_\odot$ via disk
accretion. Nowadays, there are two contending models to explain the formation of
massive stars: the core accretion model (McKee \& Tan~\cite{mckee02};
\cite{mckee03}), and the competitive accretion model (Bonnell et
al.~\cite{bonnell04}). In the core accretion model, a massive star forms from a
massive core which was fragmented from the natal molecular cloud, and gather its
mass from this massive core only. Given the non-zero angular momentum of the
collapsing core, this models predicts the existence of protostellar accretion
disks around massive stars. What is more, due to the high density in massive
cores, the accretion rate onto these disks is extremely high. On the other hand,
in the competitive accretion model, a molecular cloud initially fragments in
mainly low-mass cores, which form stars which compete to accrete mass from the
common reservoir of gas.  Therefore, massive stars should form exclusively in
clustered environments. In this scenario, the circumstellar environment is
strongly perturbed and any circumstellar disk could be severely affected and
possibly truncated by interactions with stellar companions. 

In recent years, a few Keplerian circumstellar disks and massive rotating toroids around
respectively newly born B- and O-type stars have been detected (see review by Cesaroni et
al.~\cite{cesa07} and references therein). In
particular, our group has detected a few rotating structures by searching for velocity gradients
perpendicular to molecular outflows powered by massive young stellar objects (e.g.\ Cesaroni et
al.~\cite{cesa99}; Beltr\'an et al.~\cite{beltran04}; \cite{beltran05}; Furuya et
al.~\cite{furuya08}). However, notwithstanding these important results, the controversy on massive
star formation is still open (see Beuther et al.~\cite{beuther07a} and Bonnell et
al.~\cite{bonnell07}). To pursue our investigation and establish that the disk/toroid phenomenon is a
common product of massive star formation, we have extended the sample to a larger number of objects. 
On the basis of previous experience, \MCN\ has been used as disk/toroid tracer. Note that Arce et
al.~(\cite{arce08}) and Codella et al.~(\cite{codella09}) have reported the detection of \MCN\
tracing bow shocks in the low-mass L1157-B1 molecular outflow clumps. However, in L1157, the emission
is detected far from the core harboring the embedded protostar and clearly not associated with it,
and what is more, the column densities estimated towards the L1157-B1 clumps (Codella et
al.~\cite{codella09}) are 3--4 orders of magnitude smaller than those estimated towards rotating
toroids (Beltr\'an et al.~\cite{beltran05}). In this study, we have used $^{13}$CO as molecular
outflow tracer. As mentioned above, the presence of molecular outflows has proved to be very valuable as indirect
evidence for the presence of a disk: all accretion scenarios predict the molecular outflows to be
orientated perpendicular to the accretion disks. In crowded regions, this simple picture of
perpendicularity can be more complicated due to interactions among different molecular outflows from
young stellar objects (YSOs) in the core. Therefore, when searching for rotating structures it is
very crucial to properly constrain though high-angular resolution observations the geometry and
orientation of the molecular outflows. The results of this study are presented here.

\begin{table*}
\caption[] {Parameters of the IRAM PdBI observations}
\label{par_obs}
\begin{tabular}{lcccc}
\hline
&&\multicolumn{1}{c}{Wavelength} &
\multicolumn{1}{c}{Synthesized beam$^{\rm b}$} &
\multicolumn{1}{c}{P.A.}  \\
\multicolumn{1}{c}{Source} &
\multicolumn{1}{c}{Configuration$^{\rm a}$}&
\multicolumn{1}{c}{(mm)} &
\multicolumn{1}{c}{(arcsec)} &
\multicolumn{1}{c}{(deg)}  \\
\hline
G10 &B &2.7 &$3.6\times1.6$  &$-167$  \\
&&1.4 &$2.4\times0.7$ &$-169$ \\
G19 &C &2.7 &$4.1\times1.8$ &$-161$ \\
&&1.4 &$2.6\times1.0$ &$-161$ \\ 
G29 &A,C &2.7 &$2.7\times1.3$ &$-168$ \\
&&1.4 &$1.4\times0.7$ &$-168$
\\
\hline
\end{tabular}

$^a$ Note that the A, B, and C configurations of years 2004 and 2005 are different from the
corresponding nowadays configurations.  \\
$^b$ The synthesized CLEANed beams for maps made using natural weighting.

\end{table*}

\begin{table*}
\caption[] {Frequency setups used for the molecular lines observed with the IRAM PdBI}
\label{freq_setup}
\begin{tabular}{lcccc}
\hline
&
\multicolumn{1}{c}{Center frequency} &
\multicolumn{1}{c}{Bandwidth} &
\multicolumn{2}{c}{Spectral resolution}  \\
\multicolumn{1}{c}{Line} &
\multicolumn{1}{c}{(MHz)} &
\multicolumn{1}{c}{(MHz)} &
\multicolumn{1}{c}{(MHz)}  
&\multicolumn{1}{c}{(\kms)} \\
\hline
\MCN\ (\jsc),  \MCNII\ (\jsc) &110383.508 &80 &0.156 &0.4$^{\rm a}$   \\
\MCN\ (\jsc),  \MCNII\ (\jsc) &110383.508 &320 &2.5 &6.9   \\
\MCNII\ (\jsc), $^{13}$CO (\juz) &110320.430 &160 &0.625 &1.7  \\
\MCN\  (\jsc) $v_8=1$ &110609.550 &160 &0.625 &1.7  \\
\MCN\ (\jdo) &220747.266 &80 &0.156 &0.2$^{\rm a}$   \\
\MCN\ (\jdo),  \MCNII\ (\jdo) &220679.297 &80 &0.156 &0.2$^{\rm a}$  \\
\MCN\ (\jdo),  \MCNII\ (\jdo)  &220594.422 &160 &0.625 &0.9$^{\rm b}$ \\
\MCN\ (\jdo),  \MCNII\ (\jdo),  $^{13}$CO (\jdu) &220475.812 &320 &2.5 &3.4  \\
\hline
\end{tabular}

$^a$ The spectral resolution in the maps has been degraded to 0.5~\kms. \\
$^b$ The spectral resolution in the maps has been degraded to 1.0~\kms. 

\end{table*}

\section{The sample}

We have chosen well known hot molecular cores (HMCs), whose kinematics has been studied in previous interferometric
observations, revealing velocity gradients in the cores, which are suggestive of rotation, and
evidence of outflows. Two of them lie close to an \UC\ region, and one of them contains an embedded
\UC\ region.  

{\it G10.62$-$0.38}: The G10.62$-$0.38 (hereafter G10) core, located at a distance of 3.4~kpc (Blum
et al.~\cite{blum01}) contains a well-studied \UC\ region (e.g.\ Wood \& Churchwell~\cite{wood89})
associated with the infrared source IRAS~18075$-$1956. G10 is embedded in a HMC that has been
extensively mapped in NH$_3$ (Ho \& Haschick~\cite{ho86}; Keto et al.~\cite{keto87}, \cite{keto88};
Sollins et al.~\cite{sollins05}), and more recently in SO$_2$ and OCS (Klaassen et
al.~\cite{klaassen09}). In these studies, infall and bulk rotation in the molecular gas surrounding
the \UC\ region have been detected. In addition, recent H66$\alpha$ observations have shown that
inward motions are also detected in the ionized gas (Keto ~\cite{keto02}), suggesting the existence
of an ionized accretion flow. CH$_3$OH and H$_2$O masers have been mapped towards the core and are
distributed linearly  in the plane of the rotation (Hofner \& Churchwell~\cite{hofner96}; Walsh et
al.~\cite{walsh98}), while OH masers seem to lie along the axis of rotation (Argon et
al.~\cite{argon00}). The presence of a molecular outflow  with a P.A.\  of $\sim$45$\degr$ has been
first suggested by  Keto \& Wood~(\cite{keto06}), who present a position-velocity diagram of the
high-velocity ionized jet, and Klaassen et al.~(\cite{klaassen09}), but no
high-angular resolution outflow maps are available yet. On the other hand, L\'opez-Sepulcre et al.~(\cite{lopez09})
have used the IRAM 30-m telescope to map the molecular outflow in the $^{13}$CO~(2--1) line, which shows a
direction consistent with that proposed by Keto \& Wood~(\cite{keto06}).

{\it G19.61$-$0.23}: The G19.61$-$0.23 core (hereafter G19), located at a distance of 12.6~kpc
(Kolpak et  al.~\cite{kolpak03}), is an extremely complex site associated with the infrared source
IRAS~18248$-$1158. The core contains a group of embedded \UC\ regions, firstly detected by Garay et
al.~(\cite{garay85}), and more recently mapped by Furuya et al.~(\cite{furuya05}). The G19 HMC 
has been mapped in several molecular tracers, such as CS, NH$_3$, CH$_3$CH$_2$CN, HCOOCH$_3$, and
CH$_3$CN (Plume et al.~\cite{plume92}; Garay et al~\cite{garay98}; Remijan et al.~\cite{remijan04};
Furuya et al.~\cite{furuya05}; Wu et al.~\cite{wu09}). Inverse P-Cygni profiles have been
detected in $^{13}$CO, C$^{18}$O, and CN towards the core (Wu et al.~\cite{wu09}; Furuya et
al.~\cite{furuya10}), which indicate gas infalling towards the center. A
molecular outflow has been mapped towards the core through $^{13}$CO single-dish observations
(L\'opez-Sepulcre et al.~\cite{lopez09}), but its direction is not well defined. Masers of H$_2$O,
OH, and CH$_3$OH have also been detected in this core (Forster \& Caswell~\cite{forster89}; Hofner
\& Churchwell~\cite{hofner96}; Walsh et al.~\cite{walsh98})

{\it G29.96$-$0.02}: The G29.96$-$0.02 core (hereafter G29), located at a distance of 3.5~kpc (L.\
Moscadelli, 2010, private communication) and associated with the infrared source IRAS~18434$-$0242,
contains a well-studied cometary \UC\ region (e.g.\ Cesaroni et al.~\cite{cesa94}; De Buizer et
al.~\cite{debuizer02}) with a HMC located right in front of the cometary arc (e.g.\ Wood \&
Churchwell~\cite{wood89}; Cesaroni et al.~\cite{cesa94}, \cite{cesa98}). The HMC has been mapped in
several tracers, such as NH$_3$, HCO$^+$, CS, CH$_3$CN, HNCO, HCOOCH$_3$  (Cesaroni et
al.~\cite{cesa98}; Pratap et al.~\cite{pratap99}; Maxia et al.~\cite{maxia01}; Olmi et
al.~\cite{olmi03}; Beuther et al.~\cite{beuther07b}). Cesaroni et al.~(\cite{cesa98}) and Olmi et
al.~(\cite{olmi03}) have detected in NH$_3$ and CH$_3$CN a velocity gradient across the HMC
approximately along the east-west direction, interpreted as rotation. This is similar to the velocity
gradient observed in HN$^{13}$C by Beuther et al.~(\cite{beuther07b}). A molecular outflow in the
southeast-northwest direction and centered on the HMC has been mapped in H$_2$S by Gibb et
al.~(\cite{gibb04}). The same outflow has been mapped in SiO~(8--7) by Beuther et
al.~(\cite{beuther07b}). Masers of H$_2$O, CH$_3$OH, and H$_2$CO have also been detected towards the
HMC (Hofner \& Churchwell~\cite{hofner96}; Walsh et al.~\cite{walsh98}; Hoffman et
al.~\cite{hoffman03}).

\section{Observations}

Interferometric observations of G10, G19, and G29 were carried out with the IRAM
Plateau de Bure Interferometer (PdBI) on February 28 and  March 16, 2004, and on
February 26 and March 21, 2005. G29 was observed in the most extended and
compact configurations, A and C, respectively, while cores G19 and G10 were observed in the
extended and compact ones, B and C, respectively. Due to technical problems during the
observations, for G19 the only usable configuration was the compact one, while
for G10 the only usable configuration was the extended one. Table~\ref{par_obs} reports the
configurations used for each source and the synthesized beams.
By using the dual
frequency capabilities of the PdBI we observed 
simultaneously at 2.7 and 1.4~mm.  The frequency setup of the correlator and the
list of the observed molecular lines are shown in Table~\ref{freq_setup}.
The units of the correlator were placed in
such a way that a frequency range free of lines could be used to measure the
continuum flux.
The phase centers used are indicated in Table~\ref{table_pos}. The bandpass of the receivers was calibrated by
observations of the quasar 3C273. Amplitude and phase calibrations were achieved
by monitoring 1730$-$130 and 1741$-$038, whose flux densities were determined
relative to MWC349 or 1749+096. The flux densities estimated for 1730$-$130 are
in the range 1.55--2.41~Jy at  2.7~mm, and 0.78--1.71~Jy at 1.4~mm, while those
for 1741$-$038 are in the range 3.57--3.80~Jy at 2.7~mm, and 2.18--2.77~Jy at
1.4~mm. The uncertainty in the amplitude calibration is estimated to be $\sim$20\%. The data were
calibrated and analyzed with the GILDAS\footnote{The GILDAS package is available at http://www.iram.fr/IRAMFR/GILDAS} software package developed at IRAM and
Observatoire de Grenoble. The continuum maps were created from the line free channels.
We subtracted the continuum from the line emission
directly in the {\it(u,v)}-domain. 

\begin{figure*}
\centerline{\includegraphics[angle=0,width=11.5cm]{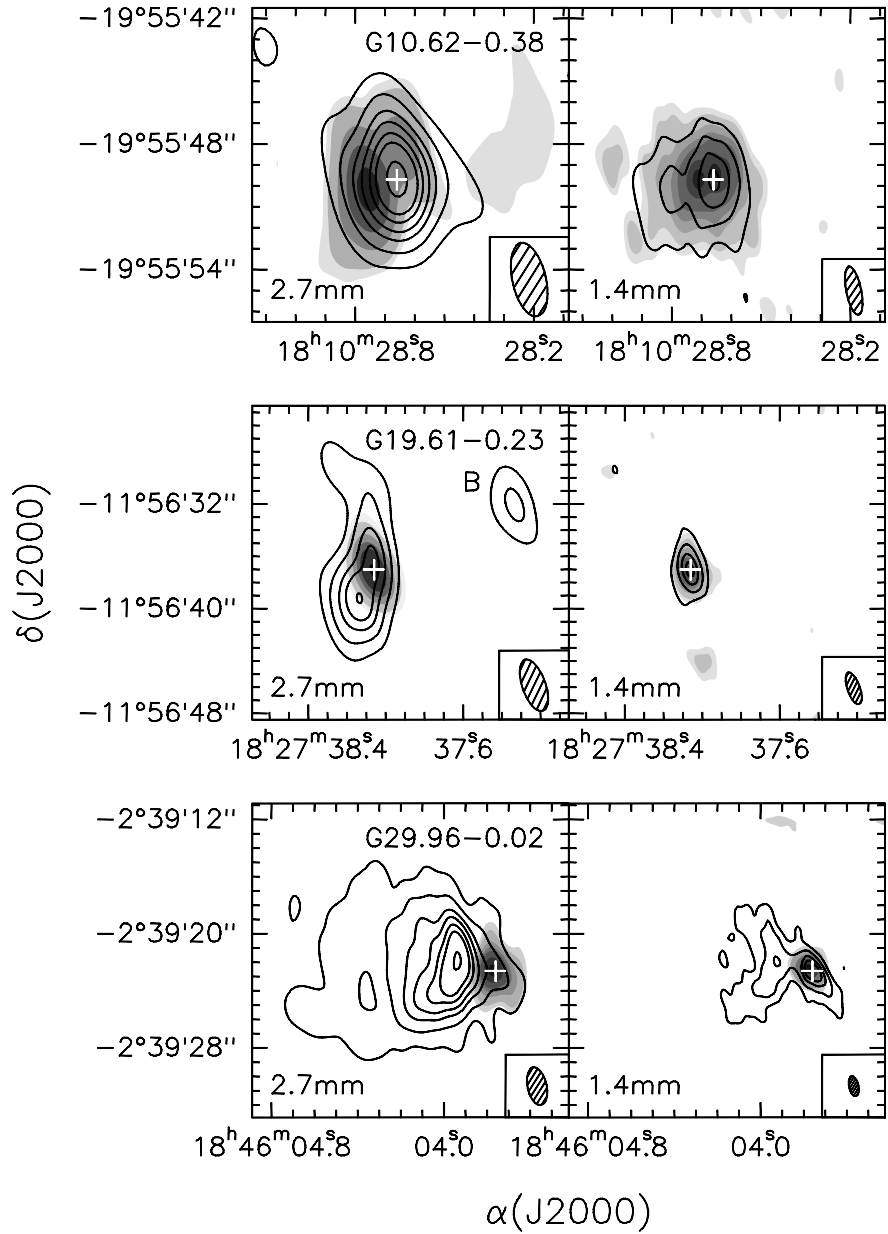}}
\caption{Overlay of the PdBI maps of the 2.7~mm ({\it left panels}) and the 1.4~mm ({\it right panels})
continuum emission ({\it contours}) on the CH$_3$CN (\jsc) emission ({\it greyscale}) averaged under the $K$ = 0, 1, 2, 3,
and 4 components ({\it left panels}) and CH$_3$CN (\jdo) emission ({\it greyscale}) averaged under the $K$ = 0, 1, 2, and 3
components ({\it right panels})  towards the cores G10.62$-$0.38, G19.61$-$0.23, and
G29.96$-$0.02.  The contour levels are 3, 9, 18, 27, 39, 51, and 75  times $\sigma$, where
1$\sigma$ is 17\mjy\ and 26.7\mjy\ at 2.7 and 1.4~mm, respectively, for G10.62$-$0.38, 
6\mjy\ and 33.3\mjy at 2.7 and 1.4~mm, respectively, for  G19.61$-$0.23, and 
3\mjy\ and 6.7\mjy\ at 2.7 and 1.4~mm, respectively, for G29.96$-$0.02. Greyscale levels are 
3, 5, 10, 15, 20, 30, and 40 times (36 for G10) $\sigma$, where 1$\sigma$ is 15~mJy\,beam$^{-1}$ at
2.7~mm and 15~mJy\,beam$^{-1}$ at 1.4~mm for G10, 30~mJy\,beam$^{-1}$ at
2.7~mm and 50~mJy\,beam$^{-1}$ at 1.4~mm for G19, and 15~mJy\,beam$^{-1}$ at
2.7~mm and 65~mJy\,beam$^{-1}$ at 1.4~mm for G29. The synthesized beam is
shown in the lower right-hand corner. The white cross marks the position of the 1.4~mm continuum 
emission peak. The source B seen towards G19.61$-$0.23 at 2.7~mm is the \UC\ region B observed by
Furuya et al.~(\cite{furuya05}).}
\label{g_cont}
\end{figure*}

\begin{table*}
\caption[] {Positions, flux densities, and diameters of the cores}
\label{table_cont}
\begin{tabular}{lcccccccccccc}
\hline
&&&\multicolumn{3}{c}{2.7~mm} &
&\multicolumn{3}{c}{1.4~mm} \\
 \cline{4-6} 
 \cline{8-10}
&\multicolumn{2}{c}{Position$^a$}
&&\multicolumn{1}{c}{TOTAL$^b$} &
\multicolumn{1}{c}{HMC$^c$} &
&&\multicolumn{1}{c}{TOTAL$^b$} &
\multicolumn{1}{c}{HMC$^c$} &&
\multicolumn{2}{c}{Source Diameter$^d$}
\\
 \cline{2-3} 
 \cline{4-6} 
 \cline{8-10}
 \cline{12-13}
&\multicolumn{1}{c}{$\alpha({\rm J2000})$} &
\multicolumn{1}{c}{$\delta({\rm J2000})$} &
\multicolumn{1}{c}{$I^{\rm peak}_\nu$} &
\multicolumn{1}{c}{$S_\nu$} &
\multicolumn{1}{c}{$S_\nu$} &&
\multicolumn{1}{c}{$I^{\rm peak}_\nu$} &
\multicolumn{1}{c}{$S_\nu$} &
\multicolumn{1}{c}{$S_\nu$} &&
\multicolumn{1}{c}{$\theta_s$} &
\multicolumn{1}{c}{$\theta_s$} \\
\multicolumn{1}{c}{Core} &
\multicolumn{1}{c}{h m s}&
\multicolumn{1}{c}{$\degr$ $\arcmin$ $\arcsec$} &
\multicolumn{1}{c}{(Jy/beam)} & 
\multicolumn{1}{c}{(Jy)} &
\multicolumn{1}{c}{(Jy)} &&
\multicolumn{1}{c}{(Jy/beam)} & 
\multicolumn{1}{c}{(Jy)} &
\multicolumn{1}{c}{(Jy)} &&
\multicolumn{1}{c}{(arcsec)} &
\multicolumn{1}{c}{(AU)} \\
\hline
G10      &18 10 28.66 &$-$19 55 49.7  &1.44   &2.42  &0.28 &&0.77 &3.40 &1.40 &&2.0 &6800 \\
G19$^e$  &18 27 38.06 &$-$11 56 37.0  &0.24   &0.64  &0.16 &&1.02 &1.50 &1.50 &&1.0 &12600 \\
G19B$^f$ &18 27 37.33 &$-$11 56 32.1  &0.07   &0.07  &... &&...$^g$  &...$^g$  &...$^g$ &&...$^g$&...$^g$ \\ 
G29      &18 46 03.76 &$-$02 39 22.6  &0.31   &2.00  &...  &&0.26 &2.03 &0.83 &&1.3 &4550  \\
\hline

\end{tabular}
   
  $^a$ Position of the 1.4~mm emission peak. \\
  $^b$ The total (free-free and dust) flux density. \\
  $^c$ Flux density of the Hot Molecular Core. See Sect.~\ref{continuum} for a description of how the HMC flux density 
  has been calculated.\\
  $^d$ Deconvolved average diameter of the 50\% contour at 1.4~mm. \\
  $^e$ The flux at 2.7~mm includes the emission of the \UC\ regions A, C, D, F, and J 
  (Furuya et al.~\cite{furuya05}).\\
  $^f$ Position and flux density at 2.7~mm of the \UC\ region B (Furuya et al.~\cite{furuya05}). \\
  $^g$ Source not detected at 1.4~mm. Position corresponds to the 2.7~mm emission peak. \\
\end{table*}

\section{Results}

\subsection{Continuum emission}
\label{continuum}

Figure~\ref{g_cont} shows the PdBI maps of the 2.7 and 1.4~mm continuum emission overlayed on the  \MCN\
(\jsc) and \MCN\ (\jdo) emission towards the three cores. The position and fluxes at 2.7 and
1.4~mm, and the deconvolved size of the sources, measured as the average diameter of the 50\% contour at
1.4~mm, are given in Table~\ref{table_cont}. As seen below, the millimeter emission at
2.7~mm of the cores is highly contaminated by free-free emission of the nearby or embedded \UC\ region(s). Due
to the fact that the 1.4~mm emission is less  affected by this problem and traces better the
HMC, the positions given in Table~\ref{table_cont} are those of the 1.4~mm emission peak. 

\subsubsection{G10.62$-$0.38}
\label{cont_g10}

The continuum dust emission of G10, which is hardly resolved at 2.7~mm, shows a compact source plus an
extended envelope. The peak of the emission at 1.4~mm coincides with that at 2.7~mm, and is very close
to the position of the \UC\ region (see Fig.~1 of Sollins et al.~\cite{sollins05}).  As seen in
Fig.~\ref{g_cont}, the continuum emission at 1.4~mm coincides with the \MCN\ (\jdo) emission.
At this wavelength, the source is clearly resolved in the east-west direction, and shows an eastern
elongation, associated with \MCN\ emission, which suggests the presence of another embedded source. On
the other hand, at 2.7~mm, the peak of the continuum emission is slightly displaced from the peak of the
\MCN\ (\jsc) averaged emission. Although we cannot rule out the possibility of the existence of
another embedded eastern source, the displacement of the line emission is likely due to the fact 
that the observations of this core have
been carried out only with the extended configuration. Therefore, part of the extended
line emission has been filtered out by the interferometer, making it very difficult to properly clean
the dirty line maps. 

To derive a rough estimate of the free-free contribution of the \UC\ region at millimeter
wavelengths, we have extrapolated the 1.3~cm emission measured by Sollins \& Ho~(\cite{sollins_ho})
assuming optically thin free-free emission ($S_\nu \propto \nu^{-0.1}$). The flux integrated density
at 1.3~cm is 2.5~Jy, and the expected free-free emission is $\sim$2.14 and $\sim$2.00~Jy at 2.7 and
1.4~mm, respectively. Therefore, the dust continuum emission associated with the HMC is 0.28 and
1.40~Jy at 2.7 and 1.4~mm, respectively (see table~\ref{table_cont}). Klaassen et
al.~(\cite{klaassen09}) have estimated a thermal dust emission of 2~Jy at 1.4~mm. However, the
free-free emission contribution at 1.4~mm estimated by these authors from radio recombination lines,
is less than half (0.92~Jy; P.\ D.\ Klaassen 2010, private communication) our estimate. The spectral
index of the dust emission between 2.7 and 1.4~mm is 2.3, which corresponds to a power-law index
$\beta$ of the dust emissivity of 0.3. This value of $\beta$ is very low and indicates that the
contribution of the free-free emission at 2.7~mm is likely higher than estimated. Therefore, 
$\beta$ should be taken as a lower limit. 




\subsubsection{G19.61$-$0.23}
\label{cont_g19}

The continuum dust emission of G19 at 2.7~mm is clearly different from that at 1.4~mm. The emission
at 2.7~mm is dominated by the free-free emission of the \UC\ regions embedded in the core
(e.g.\ Furuya et al.~\cite{furuya05}). As seen in Fig.~\ref{g_cont}, the emission shows two clumps:
the eastern and larger one associated with the \UC\ regions A, C, D, F, and J, with a morphology
similar to the one mapped at 3.5~cm by Furuya et al.~(\cite{furuya05}); and the western and smaller
one associated with the \UC\ region B. The position and flux density of this source at 2.7~mm are
reported in Table~\ref{table_cont}. On the other hand, the emission at 1.4~mm is clearly associated
with the HMC, traced by the \MCN\ (\jdo) line emission. The HMC is also visible in \MCN\ (\jsc) at
2.7~mm. This dichotomy between the emission at 2.7 and 1.4~mm is also reflected in the position of
the peak, which at 2.7~mm ($\alpha$(J2000) = 18$^{\rm h}$ 27$^{\rm m}$ 38$\fs$13,
$\delta$(J2000) = $-$11$\degr$ 56$\arcmin$ 39$\farcs$3) does not coincide with that at 1.4~mm (see
Table~\ref{table_cont}). 

Due to the complexity of the free-free emission and the number of \UC\ regions embedded in the
eastern core at 2.7~mm, it is very difficult to estimate the free-free emission contribution at
millimeter wavelengths by extrapolating the centimeter emission. Instead, we have estimated the dust
continuum emission of the HMC by measuring the flux density in a region surrounding the HMC as
traced by the \MCN~(\jsc) line. The flux density at 2.7~mm is $\sim$160~mJy, which is a value
consistent with the 147~mJy measured by Furuya et al.~(\cite{furuya05}) at 3.3~mm. At 1.4~mm, the
emission seems to be associated only with the HMC, and therefore, we have assumed that the continuum
emission is thermal dust (see Table~\ref{table_cont}). The spectral index of the HMC between 2.7 and
1.4~mm is 3.2, which corresponds to $\beta$=1.2. Furuya et al.~(\cite{furuya10}) have estimated
$\beta\gtrsim$0.7 between 3~mm and 890~$\mu$m, a value lower but still consistent with our
estimate. 

\begin{figure}
\centerline{\includegraphics[angle=0,width=7.5cm]{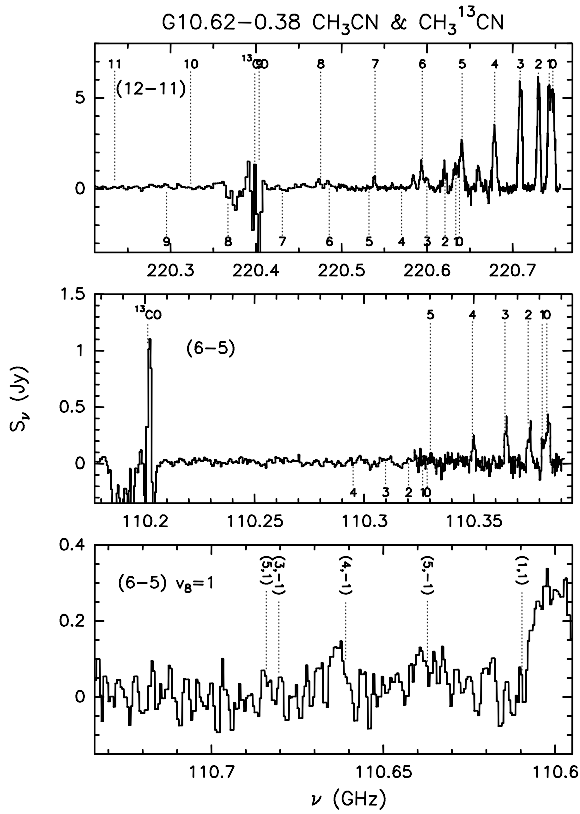}}
\caption{Methyl cyanide spectra obtained by integrating the emission inside the $3\sigma$ contour level area at 2.7 and 1.4~mm 
towards G10 as
seen with the PdBI.  We show in the top \MCN\ (\jdo), in the middle \MCN\ (\jsc),  and in the bottom
\MCN\ (\jsc) vibrationally excited ($v_8=1$). The different $K$-components ({\it top} and {\it middle panels})
are marked with dashed lines in the upper (lower) part of each spectra in the case of \MCN\ (\MCNII).
Note that different $K$-components of a same transition may have different spectral resolution 
because they were observed with different correlator units. Labeled only the analyzed lines.}
\label{g10-spectra}
\end{figure}

\begin{figure}
\centerline{\includegraphics[angle=0,width=7.5cm]{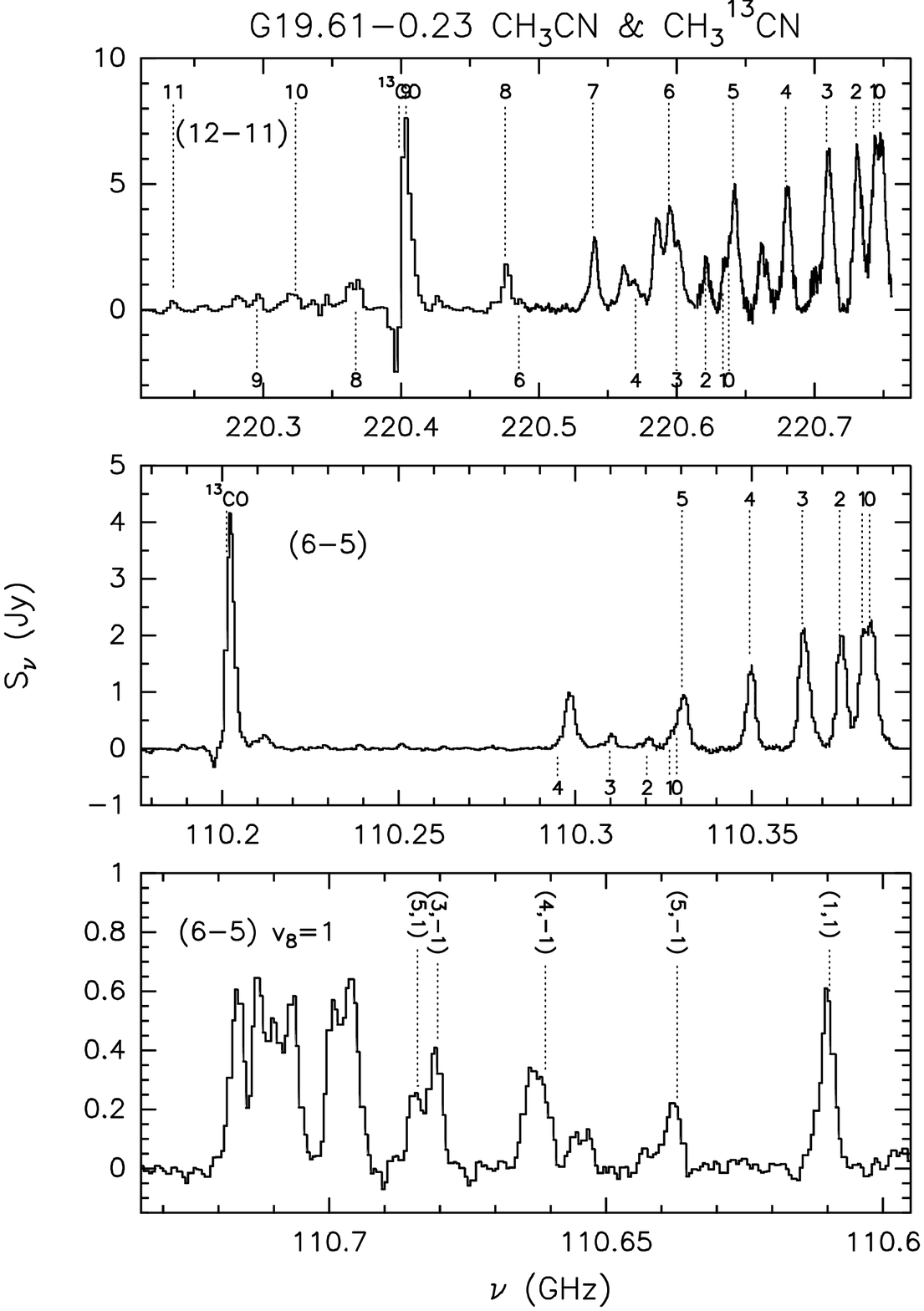}}
\caption{Same as Fig.~\ref{g10-spectra} for G19.}
\label{g19-spectra}
\end{figure}

\begin{figure}
\centerline{\includegraphics[angle=0,width=7.5cm]{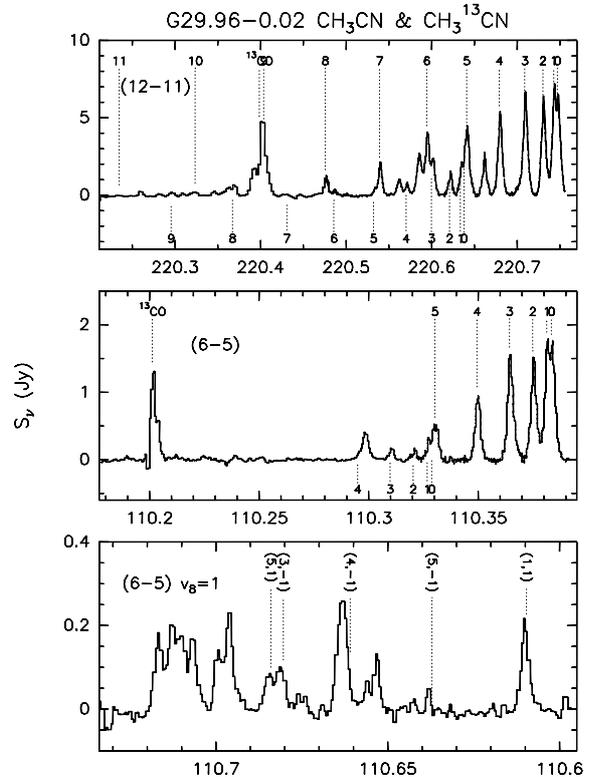}}
\caption{Same as Fig.~\ref{g10-spectra} for G29.}
\label{g29-spectra}
\end{figure}

\subsubsection{G29.96$-$0.02}
\label{cont_g29}

Alike the case of G19, also in G29 the 2.7~mm continuum emission towards the
core is different from that at 1.4~mm. At 2.7~mm, the emission is clearly
contaminated by the cometary \UC\ region and outlines the cometary arc seen in
previous observations at centimeter (e.g.\ Cesaroni et al.~\cite{cesa94}) and
mid-infrared wavelengths (De Buizer et al.~\cite{debuizer02}). Some emission is
also visible in front of the arc, associated with the HMC traced by the \MCN\
(\jsc) line  and H$_2$O maser emission (Hofner et al.~\cite{hofner96}; Olmi et
al.~\cite{olmi03}). At 1.4~mm, the cometary arc is still visible, but the
strongest continuum emission is associated with the HMC traced by the \MCN\
(\jdo) line emission in front of the arc. In fact, the peak at 2.7~mm 
($\alpha$(J2000) = 18$^{\rm h}$ 46$^{\rm m}$ 03$\fs$94, $\delta$(J2000) =
$-$02$\degr$ 39$\arcmin$ 21$\farcs$9) is shifted by $2\farcs7$ eastwards with
respect to that at 1.4~mm (Table~\ref{table_cont}).

\begin{figure*}
\centerline{\includegraphics[angle=270,width=17cm]{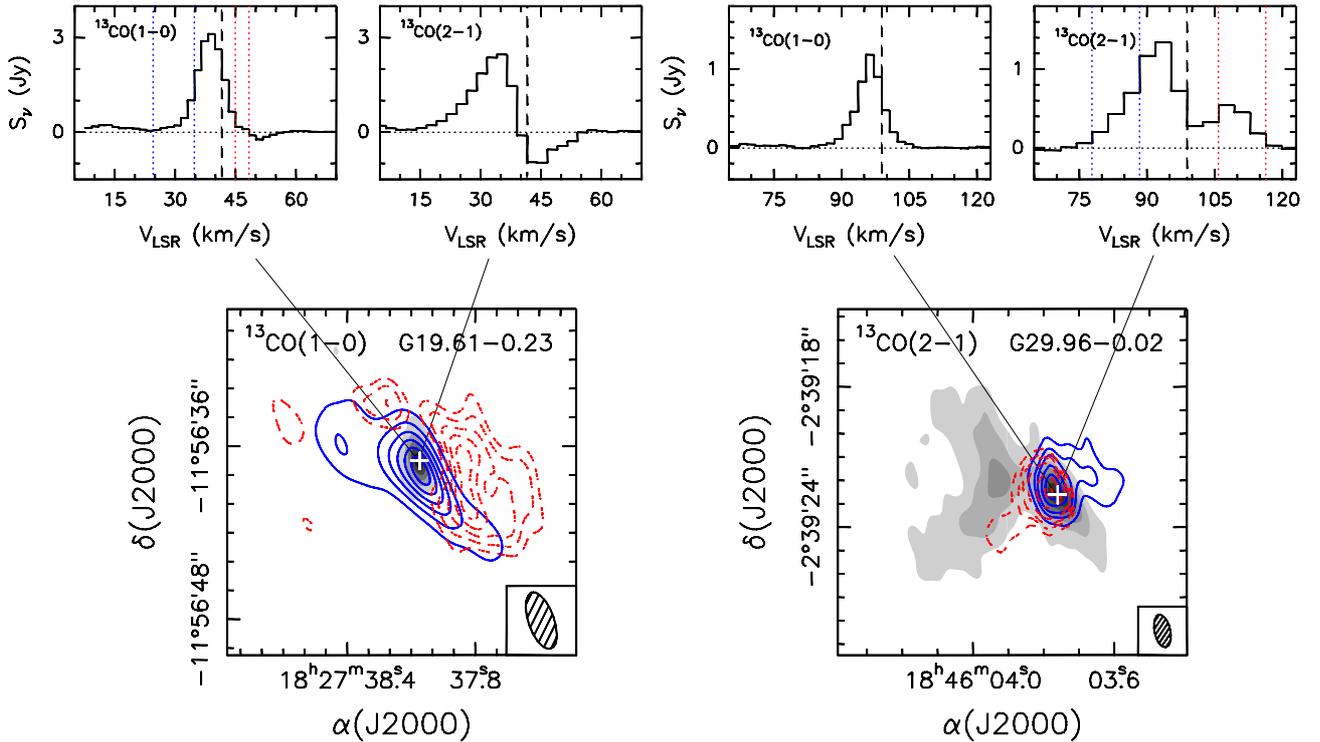}}
\caption{{\it Bottom panels}: Blueshifted ({\it blue solid contours}) and redshifted ({\it red
dashed contours})   $^{13}$CO~(\juz) ({\it left}) and $^{13}$CO~(\jdu) ({\it right}) averaged
emission overlaid on the 1.4~mm continuum emission ({\it greyscale}) towards G19 and G29,
respectively. The velocity intervals for which the blueshifted and redshifted emission have been
averaged are indicated in blue and red dotted vertical lines in the corresponding upper spectra.
Contour levels are 0.05, 0.1, 0.15, 0.2, 0.25, 0.3, and 0.35~Jy\,beam$^{-1}$
(1$\sigma\simeq$7~mJy\,beam$^{-1}$) for the $^{13}$CO~(\juz) blueshifted emission  and 0.15,
0.2, 0.25, 0.3, and 0.35~Jy\,beam$^{-1}$ (1$\sigma\simeq$50~mJy\,beam$^{-1}$) for the
$^{13}$CO~(\juz) redshifted emission  (G19), and 0.03, 0.09, 0.15, 0.27, 0.39, and
0.51~Jy\,beam$^{-1}$ (1$\sigma\simeq$7~mJy\,beam$^{-1}$) for the $^{13}$CO~(\jdu) blueshifted emission and 0.03, 0.06, 0.09, 0.15,
0.3 and 0.45~Jy\,beam$^{-1}$ (1$\sigma\simeq$7~mJy\,beam$^{-1}$) for the $^{13}$CO~(\jdu) redshifted emission (G29). Greyscale
contours for the continuum emission are the same as in Fig.~\ref{g_cont}. The synthesized beam
is shown in the lower right-hand corner. The white cross marks the position of the 1.4~mm
continuum emission peak. {\it Top panels}: $^{13}$CO~(\juz) and (\jdu) spectra taken towards the
1.4~mm continuum emission peak of the G19 ({\it left}) and G29 ({\it right}) cores. The dashed
vertical line indicates the systemic  velocity of each core.} 
\label{outflows}
\end{figure*}

To obtain a rough estimate of the free-free contribution to the 2.7~mm continuum emission, we have
extrapolated the 1.3~cm emission measured by Cesaroni et al.~(\cite{cesa94}) to 2.7~mm assuming
optically thin free-free emission. We have measured the flux density at 1.3~cm inside a  region that
corresponds to the 3$\sigma$ contour level of the 2.7~mm continuum emission and obtained a value of
2.74~Jy. Therefore, the expected free-free emission at 2.7~mm is $\sim$2.35~Jy.  This value is
higher than  the flux density measured at 2.7~mm (see Table~\ref{table_cont}). Given the fact that
the  uv-coverages of the centimeter and millimeter observations are different, it is not surprising
that the expected free-free emission value is higher than that actually measured. It is evident that
the emission at this wavelength is dominated by free-free emission. According to Olmi  et
al.~(\cite{olmi03}), who have estimated the contribution of the free-free emission at 2.7~mm by
using the same uv-coverage and restoring beam at centimeter and millimeter wavelengths, the expected
thermal dust contribution is $\sim$0.15~Jy. Using the same method, Maxia et al.~(\cite{maxia01})
estimated a dust emission of 0.24~Jy at 3.4~mm, which shows the uncertainty of these measurements.
At 1.4~mm, the emission from the HMC is clearly distinguishable from that of the cometary arc
associated with the \UC\ region. Therefore, the continuum dust emission of the HMC has been
estimated by measuring the flux density in a region surrounding it. The flux density of 0.83~Jy is
slightly higher than the value of 0.56~Jy estimated at 1.3~mm by Maxia et al.~(\cite{maxia01}) after
subtracting the free-free contribution. Maxia et al.~(\cite{maxia01}) mention that their flux
density measurement at 1.3~mm is lower than expected, and could be affected by spatial filtering of
extended emission. Adopting the values of Olmi et al.~(\cite{olmi03}) at 2.7~mm and our estimate at
1.4~mm, one obtains a spectral index of 2.5, which corresponds to $\beta$=0.5. This value of $\beta$
is very low and indicates that the contribution of the free-free emission at 2.7~mm is likely higher
than estimated. Therefore,  $\beta$ should be taken as a lower limit.



\subsection{\MCN}
 
Maps of the \MCN\ (\jsc) emission averaged under the $K$ = 0, 1, 2, 3, and 4 components, and  \MCN\
(\jdo) emission averaged under the $K$ = 0, 1, 2, and 3 components towards G10, G19, and G29 are
shown in Fig.~\ref{g_cont}. Figures~\ref{g10-spectra}, \ref{g19-spectra}, and \ref{g29-spectra} show
the spectra of the \MCN\ and \MCNII\ (\jdo) and (\jsc) lines towards the three cores. The first
vibrational state above the ground of the \MCN\ (\jsc) line, which is denoted as $v_8=1$, is also
shown in the figures. The spectra have been obtained by averaging the emission inside the $3\sigma$
contour levels of the corresponding maps in Fig.~\ref{g_cont}.

Several $K$-components of the different rotational transitions of \MCN\  and \MCNII\ at 1.4 and
2.7~mm are clearly detected towards the cores. Towards G10, only the  $K$ = 0--4 components of \MCN\
(\jsc)  are visible. Several lines of \MCN~(\jsc) $v_8=1$ are also clearly detected towards G19 and
G29 (Fig.~\ref{g10-spectra}), whereas towards G10, only few weak \MCN\
(\jsc) $v_8=1$ lines are visible. As already mentioned in Sect.~\ref{cont_g10}, the line
emission of G10 at 2.7~mm could be severely affected by the fact that the extended emission is
resolved out.

As seen in Fig.~\ref{g_cont}, the \MCN~(\jdo) emission is barely resolved in G19 and G29, and peaks at
the same position as the 1.4~mm continuum emission. On the other hand, in G10, the \MCN~(\jdo)
emission is clearly resolved and shows two peaks: a western one associated with the 1.4~mm
continuum emission peak plus a weaker one to the east.


\begin{figure}
\centerline{\includegraphics[width=8cm]{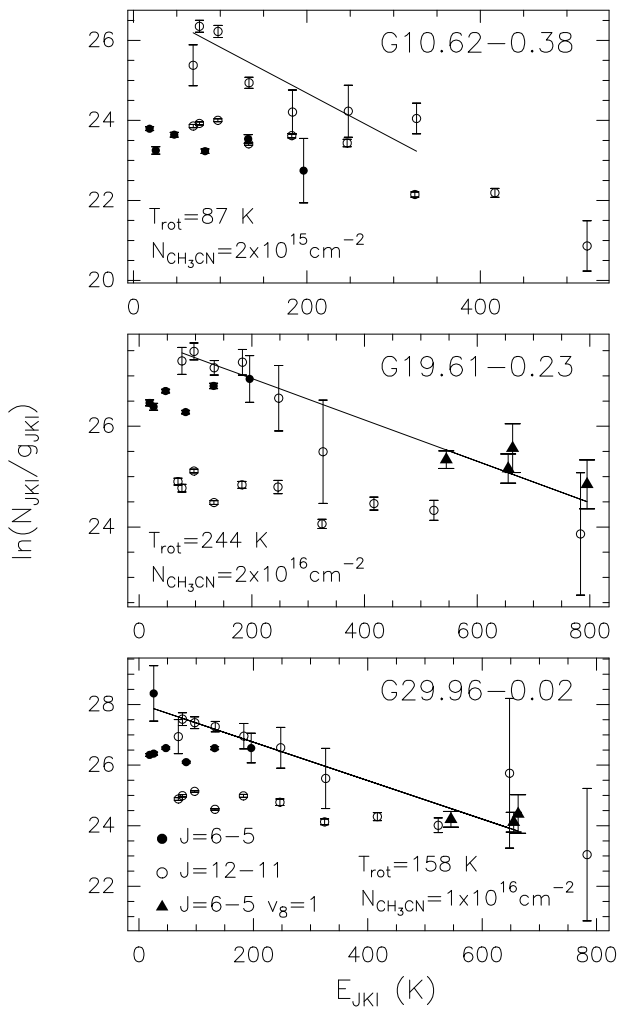}}
\vspace{0.3cm}
\caption{Rotation diagram for G10, G19, and G29 with superimposed fit. Only the
data for the
\MCNII\ transitions and the \MCN\ (\jsc) $v_8=1$ transition have been fitted.
Filled circles, open circles, and  filled triangles correspond to the 
\MCNII\ (\jsc), the \MCNII\ (\jdo), and the \MCN\ (\jsc) $v_8=1$ transition, respectively.}
\label{boltz}
\end{figure}

\subsection{$^{13}$CO}
\label{13co}

We have clearly detected the $^{13}$CO~(\juz)  and (\jdu) lines towards the cores (see
Figs.~\ref{g10-spectra}, \ref{g19-spectra}, and \ref{g29-spectra}).  In G10, the line profiles exhibit a
strong lack of emission at the center, probably due to a missing flux problem since the
in\-ter\-fe\-ro\-me\-ter has filtered out part of the extended emission. Therefore, for this core
the $^{13}$CO emission has not been analyzed.

Figure~\ref{outflows} shows the $^{13}$CO~(\juz) and (\jdu) spectra at the 1.4~mm continuum emission
peak in G19 and G29. As seen in this figure, the $^{13}$CO spectra for these two cores are also
affected by lack of emission at the central velocities. Although, we cannot exclude the possibility
that the lack of emission is due to a missing flux problem, the fact that it is observed at
redshifted velocities suggests that it could be due to absorption. This possibility will be
discussed in Sects.~\ref{g19} and \ref{g29}. In the bottom panels of Fig.~\ref{outflows}, we
present the maps of the $^{13}$CO~(\juz) averaged blueshifted and redshifted emission with respect
to the systemic velocity of G19 (41.6~\kms) and of the $^{13}$CO~(\jdu)  averaged blueshifted and
redshifted emission with respect to the systemic velocity of G29 (98.9~\kms) overlaid on the 1.4~mm
continuum emission of each core. The $^{13}$CO emission traces  a molecular outflow in each core:
one oriented east-west in G19, and one oriented southeast-northwest in G29. The latter one had been
previously mapped in H$_2$S and SiO by Gibb et al.~(\cite{gibb04}) and Beuther et
al.~(\cite{beuther07b}), respectively. Taking into account the blueshifted and
redshifted emission, the total extent of the outflows is $\sim$14$''$ ($\sim$0.86~pc) in G19, and
$\sim$6$''$ ($\sim$0.10~pc) in G29.

\section{Discussion}
\label{discuss}

\subsection{Temperature and mass estimates}
\label{rot}

CH$_3$CN is a symmetric top molecule, and therefore, can be used to estimate the
temperature in the cores. The rotational temperature, $T_{\rm rot}$, and the
total methyl cyanide column density, $N_{\rm CH_3CN}$, can be estimated by means
of the rotation diagram method, which assumes that the molecular levels are
populated according to LTE conditions at a single temperature $T_{\rm rot}$. In
the high density limit where level populations are thermalized, one expects that
$T_{\rm rot} = T_{\rm kin}$, the kinetic temperature. The CH$_3$CN ground level
transitions appear to be optically thick for all cores, as suggested by the
ratio between the main species and isotopomer. Therefore, although in the
Boltzmann plot we show all the measurements (see Fig.~\ref{boltz}), the fit was
performed only using the CH$_3^{13}$CN and vibrational excited CH$_3$CN
transitions. All the spectra were obtained by averaging the emission inside the
3$\sigma$ contour levels of the corresponding maps in Fig.~\ref{g_cont}. The
values obtained at 1.4~mm were afterwards corrected to take into account the
smaller area used to average the emission as compared to the 2.7~mm emission.
The relative abundances [CH$_3$CN]/[CH$_3^{13}$CN] were estimated following
Wilson \& Rood~(\cite{wilson94}), who give interstellar abundances as a
function of Galactocentric distance, and are  47, 44, and  50 for G10, G19, and
G29, respectively.

Figure~\ref{boltz} shows the rotational diagram together with the fits and the estimated $T_{\rm
rot}$ and $N_{\rm CH_3CN}$ for each core. The rotational temperatures obtained with the fits are 87,
244, and 158~K for G10, G19, and G29, respectively, while $N_{\rm CH_3CN}$ are 2$\times$10$^{15}$,
2$\times$10$^{15}$, and 1$\times$10$^{16}$~cm$^{-2}$ for G10, G19, and G29, respectively. Klaassen
et al.~(\cite{klaassen09}) have derived a temperature of 323$\pm$105~K for G10 using the rotational
diagram method. This value is 4 times higher than our estimated temperature. However, these authors
have used the optically thick CH$_3$CN (12--11) lines to derive $T_{\rm rot}$, and thus, their value
should be taken as an upper limit.  Wu et al.~(\cite{wu09}) have estimated a temperature of 552~K
for G19 using the rotational diagram method for CH$_3$CN~(18--17). This value is about 2.5 times
higher than our estimated temperature. This discrepancy is due to the fact that these authors have
used the optically thick CH$_3$CN lines in their calculations. In fact, Furuya et
al.~(\cite{furuya10}), who have only fitted the optically thin CH$_3^{13}$CN~(18--17) lines in the
rotation diagram, have obtained a temperature estimate of 208~K, similar to our value. Finally,
Olmi et al.~(\cite{olmi03}) have estimated a temperature of 150~K for G29 by using the rotational
diagram for CH$_3^{13}$CN~(6--5) and vibrationally excited CH$_3$CN~(6--5).This value of $T_{\rm
rot}$ is consistent with our estimate.

\begin{table*}
\caption[] {Parameters of the toroids}
\label{table_rot}
\begin{tabular}{lcccccccccccc}
\hline
&\multicolumn{1}{c}{$T_{\rm rot} ^a$} &
\multicolumn{1}{c}{$R^b$} &
\multicolumn{1}{c}{$M_{\rm gas} ^b$} &
\multicolumn{1}{c}{$V_{\rm rot}$} &
\multicolumn{1}{c}{Vel. gradient} &
\multicolumn{1}{c}{$\Delta V ^c$} &
\multicolumn{1}{c}{$\theta_{\rm CH_3CN} ^d$} &
\multicolumn{1}{c}{$M_{\rm vir} ^e$} &
\multicolumn{1}{c}{$M_{\rm dyn} ^f$} &
\multicolumn{1}{c}{$\dot M_{\rm inf} ^g$} &
\multicolumn{1}{c}{$t_{\rm inf}$} &
\multicolumn{1}{c}{$t_{\rm ff}$} \\
\multicolumn{1}{c}{Core} &
\multicolumn{1}{c}{(K)}&
\multicolumn{1}{c}{(pc)} &
\multicolumn{1}{c}{($M_\odot$)} &
\multicolumn{1}{c}{(\kms)} & 
\multicolumn{1}{c}{(\kms\,pc$^{-1}$)} & 
\multicolumn{1}{c}{(\kms)} & 
\multicolumn{1}{c}{(arcsec)} &
\multicolumn{1}{c}{($M_\odot$)} & 
\multicolumn{1}{c}{($M_\odot$)} & 
\multicolumn{1}{c}{($M_\odot$\,yr$^{-1}$)} &
\multicolumn{1}{c}{(yr)} &
\multicolumn{1}{c}{(yr)} 
\\
\hline
G10  &87   &0.016 &82   &2.1 &131 &6  &$3\farcs0$  &149 &33 &$2\times10^{-2}$  &$4\times10^3$  &$4\times10^3$  \\
G19  &244  &0.031 &415  &1.0 &32  &10 &$1\farcs2$  &615 &14 &$3\times10^{-2}$  &$15\times10^3$ &$5\times10^3$ \\
G29  &158  &0.011 &28   &1.6 &145 &9  &$1\farcs5$  &173 &13 &$8\times10^{-3}$  &$4\times10^3$  &$4\times10^3$  \\
\hline

\end{tabular}
   
  (a) From the rotational diagram method. \\ 
  (b) Radius and mass of the toroids estimated from the 1.4~mm continuum emission. \\
  (c) Line width of \MCN~(12--11). \\
  (d) Deconvolved average diameter of the 50\% contour of the \MCN~(12--11) emission. \\
  (e) Virial mass estimated assuming a spherical clump with a density distribution with $p$=1.5.
 The  values should be multiplied by 1.25 for $p$=0, and by 0.75 for $p$=2 (see
  Sect.~\ref{kine}). \\
  (f) Computed assuming an inclination angle with respect to the plane of the sky of 45$\degr$. \\
  (g) Computed assuming $V_{\rm inf}$= $V_{\rm rot}$. \\
\end{table*}

The masses of the HMCs have been estimated from the 1.4~mm dust continuum emission assuming a
dust opacity of $\simeq0.8$~cm$^{2}$\,g$^{-1}$ at 1.4~mm (Ossenkopf \& Henning~\cite{ossen94}),
a gas-to-dust ratio of 100, and a dust temperature equal to $T_{\rm rot}$. Table~\ref{table_rot}
gives the radius, $R$ and masses, $M_{\rm gas}$, of the cores together with $T_{\rm rot}$. We
derived a mass of $82\,M_\odot$ for G10. Klaassen et al.~(\cite{klaassen09}) have estimated a
mass of $136\,M_\odot$ at 1.4~mm, assuming a dust opacity of $\simeq18.87$~cm$^{2}$\,g$^{-1}$ at
2400~GHz  (Hildebrand~\cite{hildebrand83}), a dust-to-gas mass ratio of 100, $\beta$=1.5, a
temperature of 323~K (P.\ D.\ Klaassen 2010, private communication), and a distance of 6~kpc.
Using the same parameters, we obtain a mass of $102\,M_\odot$ at 1.4~mm. 

For G19, the estimated HMC mass is $415\,M_\odot$. The core mass calculated at 900~$\mu$m by Wu
et al.~(\cite{wu09}) is $15\,M_\odot$. These authors have estimated the mass by assuming
$\beta$=1.5, a temperature of 552~K and a distance of 4~kpc. Note however, that Furuya et
al.~(\cite{furuya10}) have estimated a mass of the HMC of $1300\,M_\odot$ ($1580\,M_\odot$
taking into account the other two submillimeter sources in the region) at 900~$\mu$m for a dust
opacity of $\simeq0.005$~cm$^{2}$\,g$^{-1}$ at 230~GHz (Preibisch et al.~\cite{preibisch93}), a
dust-to-gas mass ratio of 100, $\beta$=1, a temperature of 80~K, and a distance of 12.6~kpc.
Using the same dust opacity index, temperature and distance as Wu et al.~(\cite{wu09}), the
estimated dust mass would be $16\,M_\odot$, whereas the mass would be  $2142\,M_\odot$ using the
same parameters as those of Furuya et al.~(\cite{furuya10}).  On the other hand, Furuya et
al.~(\cite{furuya05}) have estimated a mass of $800\,M_\odot$ from the dust emission at 3.3~mm,
assuming a Preibisch et al.~(\cite{preibisch93}) dust opacity law, $\beta$=1.5, a temperature of
65~K, and a distance of 3.5~kpc. Using the same dust opacity law, temperature, and distance, we
would obtain a mass of 245$\,M_\odot$ from the HMC dust emission at 2.7~mm.  These discrepancies
among the values give an idea of how uncertain the values of the dust masses are and how
important to have an accurate estimate of the distance is.

For G29, we derived a HMC mass of $28\,M_\odot$. Maxia et al.~(\cite{maxia01}) have estimated a
mass of $2900\,M_\odot$ at 3.4~mm, assuming  a Preibisch et al.~(\cite{preibisch93}) dust
opacity law, a dust-to-gas mass ratio of 100, $\beta$=2, a temperature of 83~K and a distance of
6~kpc. Using the same opacity law and $\beta$ coefficient, and distance, Olmi et
al.~(\cite{olmi03}) have estimated a mass of $320\,M_\odot$ at 2.7~mm.  This could indicate that
the thermal dust flux estimated by Maxia et al.~(\cite{maxia01}) could still have some free-free
contribution. Assuming the same opacity law, temperature and distance, the gas mass estimated
from our observations at 1.4~mm would be  $270\,M_\odot$. 


\subsection{Physical parameters of the CO outflows}
\label{out}

Figure~\ref{outflows} shows the molecular outflows mapped in $^{13}$CO~(1--0) and $^{13}$CO~(2--1)
and towards G19 and G29, respectively. The parameters of the outflows are given in Table~\ref{tco}.
Note that the parameters have not been corrected for inclination. The size of the lobes $R$, mass
$M$, outflow mass loss rate $\dot M_{\rm out}$, momentum $P$, kinetic energy $E$, momentum rate in the
outflows $F$, and dynamical timescale $t_{\rm out}$ were derived from the $^{13}$CO emission for the
velocities ranges indicated in Fig.~\ref{outflows} and in Table~\ref{tco}.  The dynamical timescale
of the blueshifted and redshifted lobes has been estimated as $t_{\rm out}=R/V_{\rm out}$, where
$V_{\rm out}$ is the difference in absolute value between the maximum blueshifted or redshifted
velocity and the systemic velocity (see Table~\ref{table_pos}). The $t_{\rm out}$ of the outflow is
the maximum dynamical timescale of the two lobes.  The rest of the parameters have been calculated
for the blueshifted and redshifted lobes separately, and then added to obtain the total value. The
[$^{13}$CO]/[H$_2$] abundance ratio was estimated following Wilson and Rood~(\cite{wilson94}), and
assuming an [H$_2$]/[CO] abundance ratio of 10$^4$ (e.g.\ Scoville et al.~\cite{scoville86}). We
assumed an excitation temperature, $T_{ex}$, of 40~K. The values estimated for the G19 outflow 
are 1.7 (1.9) times smaller (higher) if $T_{ex}$ is 20~K (80~K). On the other hand, the
parameters of the G29 outflow are 1.4 (1.7) times smaller (higher) if $T_{ex}$ is 20~K
(80~K). 

As seen in Table~\ref{tco}, the values of $M$, $P$, and $E$ of the G19 outflow are about 30--90 times
higher than those of the G29 outflow. The values of $\dot M_{\rm out}$ and $F$ are comparable. The
dynamical timescale of the G29 outflow is one order of magnitude smaller than that of the G19 outflow.
The range of values obtained for G19 and G29, especially the outflow masses, are consistent with
those estimated through interferometric observations for other well-studied HMCs (e.g.\ Furuya et
al.~\cite{furuya02}, \cite{furuya08}). Regarding the mass loss rates, $10^{-4}$--$10^{-3}$~$M_\odot$
yr$^{-1}$, the values are consistent with those found for other massive
molecular outflows. Finally, the dynamical timescales of the outflows, $\sim$$10^4$~yr, are consistent
with the values estimated by Furuya et al.~(\cite{furuya02}, \cite{furuya08}).

\begin{table*}
\caption[] {Properties of the molecular outflows}
\label{tco}
\begin{tabular}{llcccccccc}
\hline
&&\multicolumn{1}{c}{$V^{a}$}&
\multicolumn{1}{c}{$R^{b}$}&
\multicolumn{1}{c}{$M^{c}$}&
\multicolumn{1}{c}{$\dot M_{\rm out}$}&
\multicolumn{1}{c}{$P^{d}$}&
\multicolumn{1}{c}{$E^{d}$}&
\multicolumn{1}{c}{$F_{\rm out} ^{d}$}&
\multicolumn{1}{c}{$t_{\rm out}$}
\\
\multicolumn{1}{c}{HMC}&
\multicolumn{1}{c}{Lobe}&
\multicolumn{1}{c}{(\kms)}&
\multicolumn{1}{c}{(pc)}&
\multicolumn{1}{c}{($M_\odot$)}&
\multicolumn{1}{c}{($M_\odot$ yr$^{-1}$)}&
\multicolumn{1}{c}{($M_\odot$ \kms)}&
\multicolumn{1}{c}{($10^{46}$ erg)}&
\multicolumn{1}{c}{($M_\odot \, \mbox{km s}^{-1}$\,yr$^{-1}$)}&
\multicolumn{1}{c}{(yr)} \\
\hline
    &Blue &[$+24.6, +34.8$]  &0.48   &78  &$2.8\times10^{-3}$ &759  &7.7  &$2.8\times10^{-2}$  &$3\times10^4$ \\
G19 &Red &[$+45, +48.4$]     &0.38   &59  &$1.1\times10^{-3}$ &298  &1.5  &$0.6\times10^{-2}$  &$5\times10^4$  \\
    &Total &                 &0.86   &137 &$2.5\times10^{-3}$ &1057 &9.2  &$2.0\times10^{-2}$  &$5\times10^4$  \\
\hline
    &Blue &[$-77.9, +88.4$]  &0.05   &0.8 &$0.4\times10^{-3}$  &12  &0.2  &$0.6\times10^{-2}$  &$2\times10^3$  \\
G29 &Red &[$+105.9, +116.4$] &0.05   &0.7 &$0.2\times10^{-3}$  &8   &0.1  &$0.3\times10^{-2}$  &$3\times10^3$	\\
    &Total &                 &0.10   &1.5 &$0.5\times10^{-3}$  &20  &0.3  &$0.7\times10^{-2}$  &$3\times10^3$  \\
\hline
\end{tabular}
 
(a) Range of outflow velocities.  \\
(b) Size of the lobe. \\		
(c) Assuming an excitation temperature of 40~K. \\
(d) Momenta and kinetic energies are calculated relative to the cloud velocity.  \\
\end{table*}

\subsection{Dense gas kinematics}
\label{kine}

In the last years, we have detected clear velocity gradients in more than 5 HMCs (Beltr\'an et
al.~\cite{beltran04}, \cite{beltran05}; Furuya et al.~\cite{furuya08}) that have been interpreted,
in most of the cases, as produced by rotating toroids oriented perpendicularly to the bipolar outflow
driven by the massive YSO(s) embedded in the HMC. As already done to search for velocity gradients in the
HMCs G24.78+0.08 and G31.41+0.31  (Beltr\'an et al.~\cite{beltran04}, \cite{beltran05}), we
simultaneously fitted multiple \MCN~(12--11) $K$-components, assuming identical line widths and fixing
their separations to the laboratory values,  at each position were \MCN\ is detected. In this case, we
fitted simultaneously the $K$ = 0, 1, 2, and 3 components.

Figure~\ref{velgrad} shows the \MCN~(\jdo) emission averaged under the $K$ = 0, 1, 2, and 3
components overlaid on the \MCN~(\jdo) line peak velocity obtained with the multiple Gaussian fits.
As seen in the LSR velocity maps (Fig.~\ref{velgrad}), all cores show clear velocity gradients,
with \Vlsr\ increasing steadily along well-defined directions. As we will see in the next sections,
the most plausible explanation for the velocity gradients observed in the HMCs is rotation.
Therefore, following Beltr\'an et al.~(\cite{beltran04}, \cite{beltran05}), we have calculated the
parameters of the rotating toroids (see Table~\ref{table_rot}). The rotation velocity $V_{\rm
rot}$,  has been estimated as half the velocity range measured in the gradients.  The virial mass
$M_{\rm vir}$ was estimated from \MCN~(12--11) assuming a spherical clump with a power-law density
distribution $\rho\propto r^p$, with $p$=1.5, and neglecting contributions from the magnetic field
and surface pressure. $M_{\rm vir}$ was computed following the expression $M_{\rm
vir}=0.407\,d\,\theta_{\rm CH_3CN}\,\Delta V^2$, where the distance $d$ is in kpc, the angular
diameter, $\theta_{\rm CH_3CN}$, in arcsec, and the \MCN~(12--11) line width, $\Delta V$, in \kms.
The  values given in Table~\ref{table_rot} should be multiplied by 1.25 for a homogeneous density
distribution, $p$=0, and by 0.75 for a density distribution with $p$=2. The dynamical mass $M_{\rm
dyn}$ was computed assuming equilibrium between centrifugal and gravitational forces from the
expression $M_{\rm dyn}=V_{\rm rot}\,R/G \sin^2 i$, where $i$ is the inclination angle of the
toroid with respect to the plane of the sky, assumed to be 90$\degr$ for an edge on toroid. Because
the inclination angle of the toroids is unknown, we arbitrarily assumed $i=45\degr$. The mass
infall rate has been computed from the expression $\dot M_{\rm inf}= M_{\rm gas}\,2\,V_{\rm inf}/R$
assuming that the infall velocity $V_{\rm inf}$ is equal to the rotation velocity following Allen
et al.~(\cite{allen03}). We have estimated the lifetimes of the toroids in two different ways, one
by measuring the infall timescale $t_{\rm inf}$ as $M_{\rm gas}/\dot M_{\rm inf}$, and the other
one by estimating the free-fall time $t_{\rm ff}$. The values are given in Table~\ref{table_rot}.
The free-fall time has been estimated from the expression $t_{\rm ff}\sim3.4\times10^7/N_{\rm
H_2}^{1/2}$ yr (Eq.~[3.5] of Hartmann~\cite{hartmann98}), where $N$, the number density of
molecular hydrogen gas, is $3\,M_{\rm gas}/4\,\pi\,R^3\,2.8\,m_H$. The parameters of the rotating
structures are consistent with those estimated by Beltr\'an et al.~(\cite{beltran04}) for the
toroids in G24.78+0.08 and G31.41+0.31 (see their Table~1).

\subsubsection{G10.62$-$0.38}

\begin{figure*}
\centerline{\includegraphics[angle=270,width=17cm]{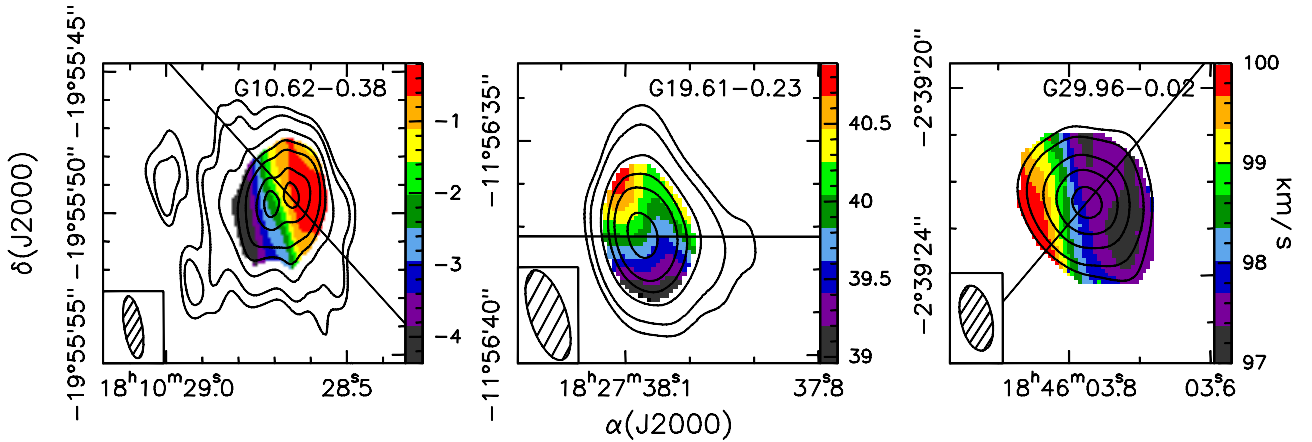}}
\caption{Overlay of the PdBI CH$_3$CN (\jdo) emission averaged under the $K$ = 0, 1, 2, and 3 components
({\it contours}) on the  CH$_3$CN (\jdo) line peak velocity obtained with a Gaussian fit ({\it colors})
towards the cores G10 ({\it left}), G19 ({\it middle}), and G29 ({\it right}). The color levels in
\kms\ are indicated in the wedge to the right of each panel. Contours for
CH$_3$CN (\jdo)
are the same as in Fig.~\ref{g_cont}. The straight line indicates the direction of the molecular
outflow (Keto \& Wood~\cite{keto06}; this paper). The synthesized beam is shown in the lower left-hand corner.}
\label{velgrad}
\end{figure*}

The \MCN\ LSR velocity map towards G10 shows a clear velocity gradient with velocity increasing
from SE to NW (Fig.~\ref{velgrad}). This velocity gradient has the same direction as that observed
in NH$_3$ by Sollins \& Ho~(\cite{sollins_ho}), and more recently in SO$_2$ by Klaassen et
al.~(\cite{klaassen09}), although, the SO$_2$ gradient is not as clear as the one seen in NH$_3$ or
in \MCN. The fact that the direction of the gradient is almost perpendicular to the direction of the
molecular outflow observed towards this HMC (see Fig.~\ref{velgrad}; Keto \& Wood~\cite{keto06}; 
Klaassen et al.~\cite{klaassen09}; L\'opez-Sepulcre et al.~\cite{lopez09}), suggests that it is
caused by the rotation of the core. 


As seen in Table~\ref{table_rot},  $M_{\rm gas}$ and $M_{\rm vir}$ are comparable, suggesting that,
in principle, turbulence could support the toroid. On the other hand,
$M_{\rm gas}$  is 2.5 times larger than $M_{\rm dyn}$, suggesting that the toroid could be unstable.
Note, however, that $M_{\rm dyn}$ has been calculated for an inclination angle $i=45\degr$ with
respect to the plane of the sky. For $i=25\degr$, the two become equal. The $\dot M_{\rm inf}$ value
is high, of the order of $10^{-2}$~$M_\odot$ yr$^{-1}$, but comparable to what has been found in
other high-mass star-forming regions (Beltr\'an et al.~\cite{beltran04}; Fontani et
al.~\cite{fontani02}). The $t_{\rm inf}$  and $t_{\rm ff}$ timescales are consistent and of the
order of a few 10$^3$~yr. Unfortunately, we have not been able to estimate the parameters of the
outflow (see Sect.~\ref{13co}) to compare the timescale of the toroid with that of the outflow, and
the infall rate with the mass loss rate.

\subsubsection{G19.61$-$0.23}
\label{g19}

The \MCN\ LSR velocity map towards G19 shows a clear velocity gradient with velocity increasing
in an almost S-N direction (Fig.~\ref{velgrad}). A similar velocity gradient has been observed in
\MCN~(18--17) by Furuya et al.~(\cite{furuya10}). As seen in Fig.~\ref{outflows}, the outflow
associated with this HMC seems to be oriented approximately E-W. Therefore, as for G10, the most
plausible explanation of this gradient is rotation.  

\begin{figure}
\centerline{\includegraphics[angle=0,width=7cm]{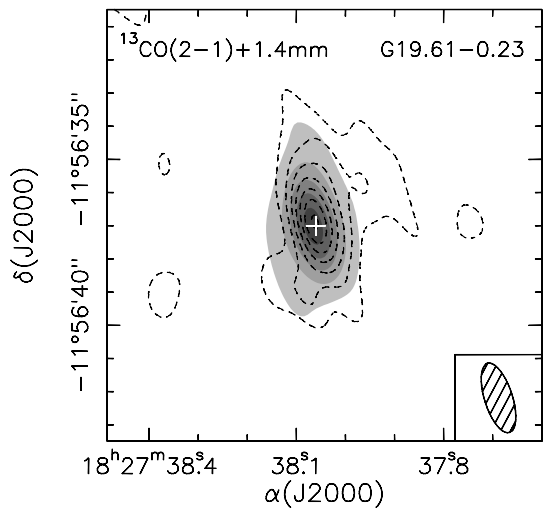}}
\caption{Overlay of the $^{13}$CO~(\jdu) averaged absorption ({\it dashed contours}) on the 1.4~mm 
continuum
emission ({\it greyscale}). The absorption has been averaged over the velocity interval 41.6--54.1~\kms.
Negative contours range from $-$0.1 to $-$0.7~Jy\,beam$^{-1}$ in steps of $-$0.1~Jy\,beam$^{-1}$
(1$\sigma\simeq$0.03~Jy\,beam$^{-1}$). Greyscale 
contours for the continuum
emission are the same as in Fig.~\ref{g_cont}. The white cross marks the position of the 1.4~mm
continuum emission peak. The
synthesized beam is shown in the lower right-hand corner.} 
\label{absorp}
\end{figure}

As seen in Table~\ref{table_rot}, $M_{\rm gas}$ and $M_{\rm vir}$ are comparable, suggesting that, in
principle, turbulence, magnetic fields, and/or thermal pressure could support the toroid. On the other hand, $M_{\rm
gas}$  is 30 times larger than $M_{\rm dyn}$. Even if $i$ were 10$\degr$, $M_{\rm gas}$ would still
be almost 2 times greater. Therefore, this suggests that the G19 toroid is highly unstable and it is
probably undergoing collapse. Further support to the collapse of this toroid is provided by
$^{13}$CO~(\jdu). As mentioned in Sect.~\ref{13co}, the $^{13}$CO~(\jdu) transition for G19 is
affected by absorption at redshifted velocities (see Fig.~\ref{outflows}). A prominent inverse
P-Cygni profile has also been observed in  $^{13}$CO and C$^{18}$O~(\jtd), and CN~(\jtd) with the
Submillimeter Array (Wu et al.~\cite{wu09}; Furuya et al.~\cite{furuya10}). The $^{13}$CO~(\jtd)
line exhibits absorption from 41.5 to 48~\kms, and the peak of the absorption is at 43.8~\kms\ (Wu
et al.~\cite{wu09}).  Such a profile is similar to the one observed by us in $^{13}$CO~(\jdu), for
which the absorption is visible from 41.6 to 54.1~\kms, and the peak is at 45.4~\kms. 
Figure~\ref{absorp} shows the map of the $^{13}$CO~(\jdu) absorption averaged over the velocity
interval 41.6--54.1~\kms\ overlayed on the 1.4~mm continuum emission. As seen in this figure, the
absorption is well correlated with the continuum emission, and the peak of the absorption coincides
with that of the continuum emission.

Similar inverse P-Cygni profiles due to redshifted absorption against the dust continuum emission or
an \HII region have already been observed towards other massive star-forming regions such as
G24.78+0.08 (Beltr\'an et al.~\cite{beltran06}), G31.41+0.31 (Girart et al.~\cite{girart09}; 
Frau et al., in preparation), and W51 IRS2 (Zapata et al.~\cite{zapata08}). In all cases, the
redshifted absorption has been interpreted as the signature of infall towards the core center.
The same interpretation has been proposed by Wu et al.~(\cite{wu09}) and Furuya et
al.~(\cite{furuya10}) for the inverse P-Cygni
profiles observed in $^{13}$CO and C$^{18}$O~(\jtd), and CN~(\jtd) towards G19. 
The $^{13}$CO~(2--1) line brightness temperature ($T_{\rm B} \simeq -10.9$~K) measured along the
line-of-sight towards the position of the HMC is comparable to the continuum brightness
temperature ($T_{\rm c} \simeq 10.3$~K). Hence, we have not been able to estimate the excitation temperature of the
absorbing gas.

Taking into account that the systemic velocity \Vlsr\ of G19 is 41.6~\kms and that the peak of the
redshifted absorption is at 45.4~\kms, we have estimated the infall velocity as $V_{\rm inf}$ =
$\vert$\Vlsr--$V_{\rm redshifted}$$\vert$ = 3.8~\kms, following Beltr\'an et al.~(\cite{beltran06}).
The $V_{\rm inf}$ estimated from  $^{13}$CO~(\jtd) is 3.5~\kms\, (Wu et al.~\cite{wu09}) and
4~\kms\, (Furuya et al.~\cite{furuya10}), whereas that from CN~(\jtd) is 6.0~\kms\, (Wu et
al.~\cite{wu09}). Note that the $V_{\rm inf}$ estimated from the redshifted absorption is higher
than the infall velocity used to estimate $\dot M_{\rm inf}$ and $t_{\rm inf}$ in
Table~\ref{table_rot}, which is 1~\kms. This value is a rough estimate of the infall velocity, hence
the discrepancy between the two $V_{\rm inf}$ estimates is not surprising. What this discrepancy
indicates is that infall is stronger than rotation. Assuming $V_{\rm inf}$ = 3.8~\kms, $\dot M_{\rm
inf}$ estimated from $M_{\rm gas}$ (see Sect.~\ref{kine}) is 0.1~$M_\odot$ yr$^{-1}$, and $t_{\rm
inf}$ = 4$\times$10$^3$~yr. This latter value is more similar to the estimated $t_{\rm ff}$.

Following Beuther et al.~(\cite{beuther02}), one can estimate the mass accretion rate $\dot M_{\rm
acc}$ from the outflow mass loss rate $\dot M_{\rm out}$. According to these authors, the mass loss
rate of the outflow is related to the mass loss rate of the internal jet entraining the outflow $\dot
M_{\rm jet}$ as $\dot M_{\rm out}$ = $\dot M_{\rm jet}\, V_{\rm jet}/V_{\rm out}$, where the ratio
between the jet velocity $V_{\rm jet}$ and the molecular outflow velocity $V_{\rm out}$ is $\sim$20.
Assuming a ratio between $\dot M_{\rm jet}$ and the mass accretion rate onto the protostar $\dot
M_{\rm acc}$ of approximately 0.3 (Tomisaka~\cite{tomisaka98}; Shu et al.~\cite{shu99}), one finds
that  $\dot M_{\rm out}$ = 20\,$\dot M_{\rm jet}$ = 6\, $\dot M_{\rm acc}$. Therefore, for G19, $\dot
M_{\rm acc}$ would be $\sim$4$\times$10$^{-4}$~$M_\odot$ yr$^{-1}$.  $\dot M_{\rm acc}$ is about two
orders of magnitude smaller than $\dot M_{\rm inf}$ (see Table~\ref{table_rot}). Note that the
outflow parameters have not been corrected for inclination. $\dot M_{\rm out}$ is to be multiplied by
$\tan i$ to correct for the inclination angle $i$ of the flow with respect to the line-of-sight.
However, a correction by a factor $>10$ would imply $i>84\degr$, which seems very unlikely because
rotation would be very difficult to detect in such a face-on structure. Therefore, the outflow
inclination cannot account for the difference between $\dot M_{\rm acc}$ and $\dot M_{\rm inf}$. 
In addition, part of the extended outflow emission could have been filtered out by the
interferometer, and therefore $\dot M_{\rm out}$ and $\dot M_{\rm acc}$ should be considered as lower
limits.  However, after inspecting the $^{13}$CO channel maps, we conclude that the missing 
flux problem is unlikely to account for 2 orders of magnitude of difference in the estimates of
the outflow parameters. Moreover, the single-dish study of L\'opez-Sepulcre et al.~(\cite{lopez10}) have obtained a similar
result with mass
accretion rates 2 to 4 orders of magnitude smaller than the infall rates for a sample of
high-mass cluster forming clumps. Their interpretation is that $\dot M_{\rm inf}$ represents the
infall of the material onto a cluster of stars, while $\dot M_{\rm acc}$ corresponds
to the material being accreted onto a single massive star, responsible for the massive molecular
outflow detected. The large masses and luminosities involved suggest that these rotating
toroids likely host a stellar cluster rather than a single star (Beltr\'an et
al.~\cite{beltran05}; Cesaroni et al.~\cite{cesa06}).

\subsubsection{G29.96$-$0.02}
\label{g29}

\begin{figure}
\centerline{\includegraphics[angle=-90,width=8cm]{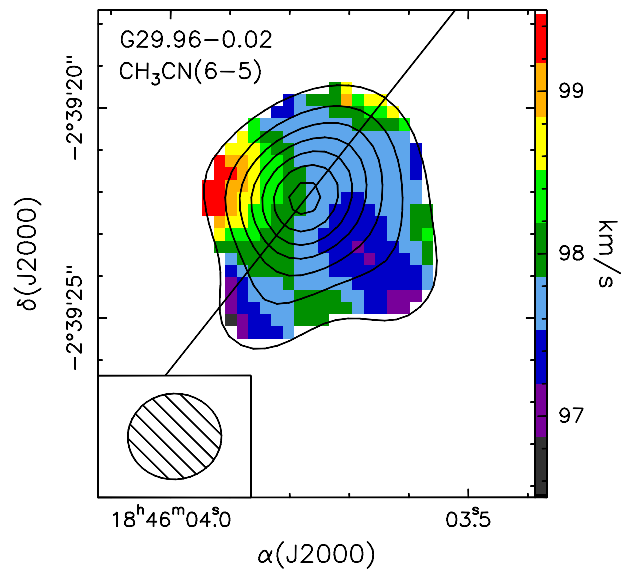}}
\caption{Overlay of the OVRO CH$_3$CN (\jsc) emission from Olmi et al.(~\cite{olmi03}) averaged under the $K$ = 0 and 1 components
({\it contours}) on the  CH$_3$CN (\jsc) line peak velocity obtained with a Gaussian fit ({\it colors})
towards the core G29. The color levels in
kilometers per second are indicated in the wedge to the right. Contours for CH$_3$CN (\jsc)
are range from 0.1 to 0.7~Jy\,beam$^{-1}$ in steps of 0.1~Jy\,beam$^{-1}$ (1$\sigma\simeq$0.03~Jy\,beam$^{-1}$).  The straight line indicates the direction of the molecular
outflow. The
synthesized beam is shown in the lower left-hand corner.} 
\label{ovro}
\end{figure}

The \MCN\ LSR velocity map towards G29 shows a clear velocity gradient with velocity increasing
in from W to E (Fig.~\ref{velgrad}). Cesaroni et al.~(\cite{cesa98}) and Olmi et al.~(\cite{olmi03})
have detected in NH$_3$ and CH$_3$CN a velocity gradient across the HMC approximately along the same
direction by fitting the peak of the emission at each spectral channel. This is similar to the
velocity gradient observed in HN$^{13}$C by Beuther et al.~(\cite{beuther07b}). As seen in
Fig.~\ref{outflows}, the embedded YSO(s) in this HMC powers a molecular outflow with a SE-NW
direction. This outflow has also been mapped in  H$_2$S (Gibb et al.~\cite{gibb04}) and SiO~(8--7)
(Beuther et al.~\cite{beuther07b}). The velocity gradient is clearly not perpendicular to the
molecular outflow. Note that also for G10 and G19 the velocity gradients are not exactly
perpendicular to the molecular outflows (see Fig.~\ref{velgrad}). Although for these HMCs the
discrepancy is of a few degrees. One possible explanation for the non-perpendicularity of the
velocity gradient could be the highly elliptical beam of the PdBI observations. According to
Guilloteau \& Dutrey~(\cite{guilloteau98}), to analyze the velocity field, it is essential to use a
circular beam, otherwise, velocity gradients in marginally resolved objects may be severely
distorted. To investigate this effect, we re-analyze the OVRO CH$_3$CN~(\jsc) observations towards
G29 carried out by Olmi et al.~(\cite{olmi03}), with the simultaneous fit of the \MCN\
$K$-components. Figure~\ref{ovro} shows the \MCN~(\jsc) emission averaged under the $K$ = 0
and 1 components overlaid on the \MCN~(\jsc) line peak velocity. As seen in this plot, the
synthesized beam of these observations is almost circular, and the velocity gradient shows a SW-NE
direction. This direction is perpendicular to that of the molecular outflow, and therefore, we
interpret the velocity gradient as due to rotation. 

As seen in Table~\ref{table_rot}, $M_{\rm vir}$ is 6 times higher than $M_{\rm gas}$, suggesting that,
in principle, turbulence, magnetic fields, and/or thermal pressure could support the toroid. On the
other hand, $M_{\rm gas}$  is 2 times larger than $M_{\rm dyn}$. For $i$$\sim$30$\degr$, the two
become comparable.
The difference between $M_{\rm gas}$ and $M_{\rm dyn}$ could suggest that the G29 toroid is
 unstable and probably undergoing collapse.  As mentioned in Sect.~\ref{13co}, the
$^{13}$CO~(\jdu) transition towards the G29 HMC is affected by absorption at redshifted velocities
(see Fig.~\ref{outflows}). In this case, the absorption is less clear than for G19. Probably this is
because emission and absorption are mixed along the line-of-sight towards the HMC, making the latter
less evident. The peak of the absorption is at $\sim$104~\kms, therefore $V_{\rm inf}$ is
$\sim$5~\kms. Note that the $V_{\rm inf}$ estimated from the redshifted absorption is higher than the
infall velocity used to estimate $\dot M_{\rm inf}$ and $t_{\rm inf}$ in Table~\ref{table_rot}, which
is 1.6~\kms. Assuming $V_{\rm inf}$ = 5~\kms, $\dot M_{\rm inf}$ estimated from $M_{\rm gas}$ (see
Sect.~\ref{kine}) is 0.03~$M_\odot$ yr$^{-1}$, and $t_{\rm inf}$ = 1$\times$10$^3$~yr. As already seen
for the other toroids, $\dot M_{\rm inf}$ is high, of the order of $10^{-2}$~$M_\odot$ yr$^{-1}$,  but
comparable to what has been found in other high-mass star-forming regions. The $t_{\rm inf}$  and
$t_{\rm ff}$ timescales are consistent and of the order of $\sim$10$^3$~yr.

The mass accretion rate $\dot M_{\rm acc}$ estimated from the outflow mass loss rate $\dot M_{\rm
out}$ (see previous section) is $\sim$8$\times$10$^{-5}$~$M_\odot$ yr$^{-1}$. As for G19 (see
previous section), the accretion rate is two orders of magnitude smaller than the infall rate (see
Table~\ref{table_rot}), and neither the inclination of the outflow nor the filtering out of the
emission seem able to account for such a discrepancy. Therefore, also in this case, the most
plausible explanation for $\dot M_{\rm inf} \gg \dot M_{\rm acc}$ is that the material of the toroid
is infalling onto a cluster of stars instead of a single star, responsible for the molecular
outflow.

\subsubsection{Absorption towards the G29.96$-$0.02 \UC\ region}

\begin{figure}
\centerline{\includegraphics[angle=0,width=7.5cm]{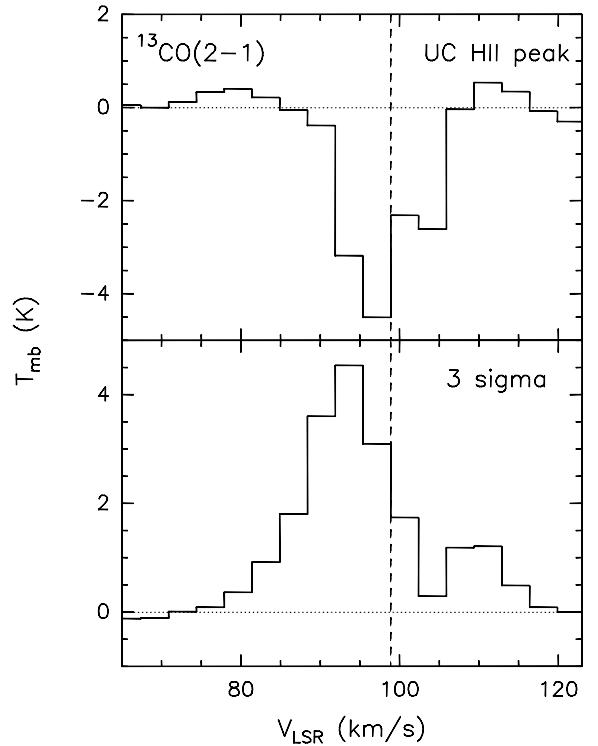}}
\caption{$^{13}$CO~(2--1) integrated spectra towards the peak of the 1.4~mm continuum emission of the \UC\
region ({\it top}), and over the 3$\sigma$ contour level of
the 1.4~mm continuum  emission ({\it bottom}) towards G29. The dashed vertical line indicates the systemic 
velocity of each core.
} 
\label{hii_abs}
\end{figure}

\begin{figure}
\centerline{\includegraphics[angle=0,width=7cm]{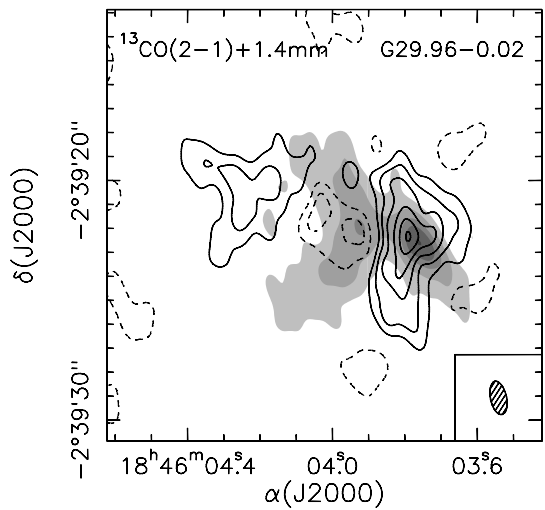}}
\caption{Overlay of the $^{13}$CO~(\jdu) mean absorption ({\it dashed contours}) on the 1.4~mm 
continuum emission ({\it greyscale}). The absorption has been averaged over the velocity interval
98.9--105.9~\kms. Negative contours are $-$0.06 and $-$0.12~Jy\,beam$^{-1}$, while positive
contours range from 0.08 to 0.56 in steps of 0.08~Jy\,beam$^{-1}$ (1$\sigma\simeq$0.02~Jy\,beam$^{-1}$). Greyscale  contours for the
continuum emission are the same as in Fig.~\ref{g_cont}.  The synthesized beam is shown in the lower right-hand corner.} 
\label{g29_abs}
\end{figure}

Maxia et al.~(\cite{maxia01}) detected redshifted absorption in the integrated spectra of
HCO$^+$~(1--0) towards G29. The map of the HCO$^+$~(1--0) line averaged over the absorption velocity
range, 99--102~\kms, shows that the absorption is positionally coincident with the \UC\ region (see
Fig.~8 of Maxia et al.~\cite{maxia01}). To check whether this absorption is also visible in the
$^{13}$CO~(2--1) line, we have obtained the spectrum towards the peak of the 1.4~mm continuum emission
of the \UC\ region, and, following Maxia et al.~(\cite{maxia01}), over the 3$\sigma$ contour level of
the 1.4~mm continuum emission towards G29; that is, a region including the HMC and the \UC\ region
(Fig.~\ref{hii_abs}). As seen in the top panel of this figure, deep and broad absorption is
clearly visible towards the \UC\ region. When integrating the $^{13}$CO~(2--1) emission over the
3$\sigma$ contour level of the 1.4~mm continuum emission, the emission combines with the absorption,
making it less pronounced. However, the absorption is still clearly visible at a velocity of
$\sim$104~\kms. This velocity is different from that of the HCO$^+$ absorption, which according to
Maxia et al.~(\cite{maxia01}) is 98.8~\kms. This may mean that this $^{13}$CO absorption feature
is not real but due to the interferometer resolving out extended emission at this velocity.
Alternatively, the $^{13}$CO emission could be stronger and more extended
than the HCO$^+$ emission, and thus, the combination of emission and absorption could produce an
absorption feature at a velocity interval different from that observed in HCO$^+$. From
Fig.~\ref{hii_abs} and the $^{13}$CO~(2--1) channel maps, one sees that the redshifted absorption is
visible towards the position of the \UC\ region in the velocity range 98.9--105.9~\kms. The
$^{13}$CO~(2--1) map averaged over this velocity interval is shown in Fig.~\ref{g29_abs}. This map is
similar to the HCO$^+$ integrated map of Maxia et al.~(\cite{maxia01}), and clearly shows absorption
towards the \UC\ region, whereas emission is visible towards the HMC and the NE. For these
reasons, we believe that the $^{13}$CO absorption feature at $\sim$104~\kms\ is real, although only
complementary single-dish data may prove this.

The $^{13}$CO~(2--1) line brightness temperature $T_{\rm B}$ measured along the
line-of-sight towards the position of the \UC\ region is $-$4.5~K. Within the uncertainties,
this is comparable to the continuum brightness
temperature $T_{\rm c} \simeq 3.4$~K, which shows that the continuum photons are almost totally
absorbed by the $^{13}$CO gas.




\section{Rotating toroids in massive star forming regions}
\label{toroids}

In recent years a number of massive, rotating structures have been discovered in high-mass YSOs (see
Table~\ref{minf}). Following Cesaroni et al.~(\cite{cesa06}), these rotating structures should be
classified into two classes on the basis of the ratio between the mass of the rotating structure and
that of the star. On the one hand, rotating structures with masses lower than the mass of the central,
star, typically found around B-type stars, are centrally supported disks. On the other hand,
structures having masses in excess of several 10~$M_\odot$, much greater than the mass of the central
star(s), and found around O-type stars, should be called toroids.

In Beltr\'an et al.~(\cite{beltran04}, \cite{beltran05}) and this work, we have concentrated on the
study of the large rotating toroids around O-type stars, deriving their properties and comparing them
to disks around low-mass YSOs. The sizes of the
toroids are of several 1000~AU, which are an order of magnitude higher than those of accretion disks
in solar-type pre-main sequence stars ($\sim$ a few 100~AU). As already discussed, their masses 
are much higher
than the mass of any central (proto)star. In fact, these toroids are so huge that they may host not
just a single star, but a whole cluster. This is completely different from the low-mass scenario,
where the disks have masses of a few 10$^{-3}$ to 10$^{-1}$~$M_\odot$ (Natta~\cite{natta00}), i.e.\
10--100 times smaller than the mass of the central star. 

For some of the HMCs hosting rotating toroids, inverse P-Cygni profiles (or redshifted absorption),
have also been detected (this work; Sollins et al.~\cite{sollins05}; Beltr\'an et
al.~\cite{beltran06}; Zapata et al.~\cite{zapata08}; Girart et al.~\cite{girart09}; Wu et
al.~\cite{wu09}; Furuya et al.~(\cite{furuya10})), indicating that these toroids are not only rotating
but also infalling. However, one sees that the infall rates are of the order of
10$^{-3}$--10$^{-2}$~$M_\odot$ yr$^{-1}$, while those derived for accretion disks in low-mass YSOs
range from 10$^{-9}$ to 10$^{-6}$~$M_\odot$ yr$^{-1}$ (Hartmann~\cite{hartmann98}). 

Based on this and on the fact that the infall rates are 2 orders of magnitudes greater than the
mass accretion rates estimated from the mass loss rates of the corresponding outflows
(Sects.~\ref{g19} and \ref{g29}), we propose an scenario in which massive toroids would be
transient structures infalling towards a central cluster of forming stars. In this scenario, the
circumcluster toroids would be fed by a larger scale reservoir of material, consisting of the
parsec-scale clumps surrounding them. The material of the toroids would infall onto the
circumstellar disks in the cluster, and then from the disks it
would accrete onto the corresponding central stars. In this sense, the circumcluster toroids
would be the high-mass analogs of the circumstellar infalling envelopes surrounding low-mass
stars, and the embedded circumstellar disks (not imaged yet towards O-type stars; Cesaroni
et al.~\cite{cesa06}, \cite{cesa07}) would correspond to the accretion disks observed
towards Class~0 and I low-mass YSOs.

\begin{figure}
\centerline{\includegraphics[angle=-90,width=9cm]{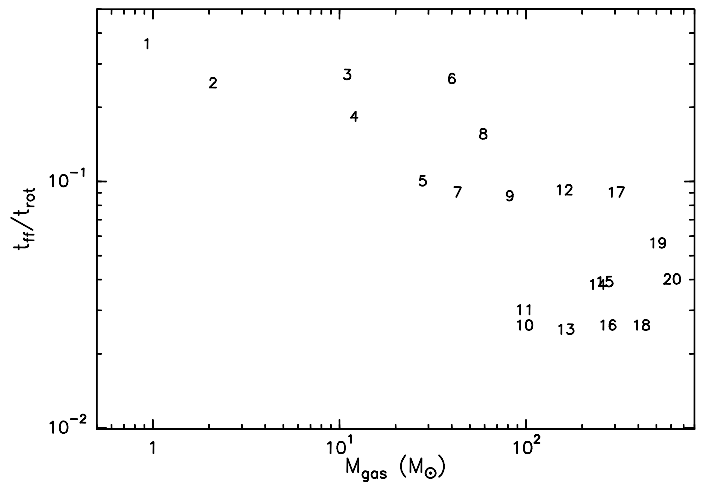}}
\caption{Free-fall timescale to rotational period ratio versus gas mass of known rotating disks or
toroids. The masses have been estimated assuming the dust opacities of Ossenkopf \&
Henning~(\cite{ossen94}). The numbers correspond to the entries of Table~\ref{minf}.} 
\label{minfall}
\end{figure}

The differences in the velocity field observed towards low- and high-mass YSOs also support this
idea that toroids are qualitatively different from disks. For low-mass stars, the disks undergo
Keplerian rotation, while for toroids, Keplerian rotation is not possible on scales of 10$^4$~AU
due to the fact that the gravitational potential of the system is dominated by the massive toroid,
not by the star. For this reason, it has been proposed that these toroids never reach equilibrium
(Cesaroni et al.~\cite{cesa06}). They could be transient entities, with timescales of the order of
the free-fall time, $\sim$10$^4$~yr. In contrast, according to Natta~(\cite{natta00}), low-mass
pre-main-sequence disks may live as long as 10$^7$~yr. To study the stability of these structures,
we have plotted in Fig.~\ref{minfall} the $t_{\rm ff}/t_{\rm rot}$ ratio versus $M_{\rm gas}$,
where the free-fall time $t_{\rm ff}$ is proportional to the dynamical timescale needed to refresh
the material of the toroid and  $t_{\rm rot}$ is the rotational  period at the outer radius,
$2\,\pi\,R/V_{\rm rot}$, for a number of disks and toroids around high-mass (proto)stars.  The gas masses found in the literature have been estimated assuming different dust opacity laws
and, as seen in Sect.~\ref{rot}, this is an important source of uncertainty in the mass
determination. Therefore, for all but one case (GH2O 92.67+3.07\footnote{The mass of this core 
has been estimated from CS emission (Bernard et al.~\cite{bernard99}).}), we have re-calculated the gas
masses from the dust continuum emission using the dust opacities of Ossenkopf \&
Henning~(\cite{ossen94}), with a dust opacity of $\simeq$0.8~cm$^2$\,g$^{-1}$ at 1.4~mm, i.e.\ the
same used for our sources. We give this in col.~3 of Table~\ref{minf}, while in col.~4 we give the
mass estimates found in the literature. Note that for those wavelengths for which the opacity has
not been tabulated by Ossenkopf \& Henning~(\cite{ossen94}), we have extrapolated the value from
$\lambda$=1.4~mm assuming $\beta$=2. These gas masses have been used to estimate $t_{\rm ff}$,
which together with $t_{\rm rot}$, have been calculated following Sect.~\ref{kine} and using the
parameters reported in Table~\ref{minf}. Assuming an uncertainty of $\sim$20\% in $M_{\rm gas}$, $V_{\rm rot}$,
and $R$, the uncertainty in $t_{\rm ff}/t_{\rm rot}$ would be $\sim$25\%. In all cases, If the structure rotates fast, the
infalling material has enough time to settle into a centrifugally supported disk. Vice versa, if
the structure rotates slowly,  the infalling material does not have enough time to reach
centrifugal equilibrium and the rotating structure is a transient toroid. Therefore, the higher
the $t_{\rm ff}/t_{\rm rot}$ ratio, the more similar should be the rotating  structure to a
circumstellar disk. In fact, in Fig.~\ref{minfall}, one sees that the less massive structures,
which have masses  comparable to or lower than that of the central star  have the higher $t_{\rm
ff}/t_{\rm rot}$ ratio, while the massive toroids have a lower ratio. A typical example of a
circumstellar disk  in Keplerian rotation around a B-type star is  that found towards
IRAS~20126+4104 (Cesaroni et al.~\cite{cesa05}), while a typical example of rotating toroid around
an O-type star is G31.41+0.31 (Beltr\'an et al.~\cite{beltran04}, \cite{beltran05}).



In summary, our findings suggest that massive stars could form by infall/accretion through
rotating toroids/disks, although the sizes, masses, timescales, infall, and accretion rates
involved are much greater than those of low-mass accretion disks. It is important to stress that
although rotation, infall, and outflow have been detected towards massive YSOs, so far no real
accretion disk in O-type stars  has been imaged yet. However, as already discussed by Cesaroni
et al.~(\cite{cesa06}), real accretion disks might be embedded inside the rotating toroids,  and
impossible to disentangle from the massive toroids with current instrumentation. Only the new
capabilities of the Atacama Large Millimeter Array (ALMA), in terms of sensitivity and
resolution, will shed some light on this important issue.

\begin{table*}
\caption[] {List of rotating disks and toroids in high-mass (proto)stars}
\label{minf}
\begin{tabular}{clcccccc}
\hline
&&\multicolumn{1}{c}{$M_{\rm gas}^{\rm OH94\,\,b}$} &
\multicolumn{1}{c}{$M_{\rm gas}^{c}$} &
\multicolumn{1}{c}{$R$} &
\multicolumn{1}{c}{$V_{\rm rot}$} 
\\
\multicolumn{1}{c}{Number$^a$} &
\multicolumn{1}{c}{Core} &
\multicolumn{1}{c}{($M_\odot$)} &
\multicolumn{1}{c}{($M_\odot$)} &
\multicolumn{1}{c}{(pc)}&
\multicolumn{1}{c}{(\kms)} &
\multicolumn{1}{c}{Refs.$^{d}$} &
\multicolumn{1}{c}{$t_{\rm ff}/t_{\rm rot}^{\,\,e}$} \\
\hline
1   &IRAS~20126+4104        &0.93     &4       &0.008	  &1.3	    &1,2     &0.36  \\
2   &Cepheus A HW2      &2.1      &8$^f$   &0.0016	  &3.0	    &3	     &0.25  \\
3   &IRAS~23151+5912        &11       &26      &0.010	  &3.0	    &4	     &0.27  \\
4   &GH2O 92.67+3.07    &12$^g$   &12      &0.035	  &1.2	    &5	     &0.18  \\
5   &G29.96$-$0.02   	&28	  &28	   &0.011         &1.6      &6       &0.10  \\
6   &G20.08$-$0.14N  	&40	  &95$^h$  &0.024         &3.5$^i$  &7       &0.26  \\
7   &NGC 6334I	     	&43	  &17	   &0.0014        &5.1      &8, 9    &0.09  \\
8   &IRAS~18089$-$1732   	&59       &45$^j$  &0.010         &4.0      &10      &0.16  \\
9   &G10.62$-$0.38   	&82	  &82	   &0.016         &2.1      &6       &0.09  \\
10  &G28.87+0.07     	&98	  &100     &0.029         &0.5      &11      &0.03  \\
11  &G24.78+0.08 C   	&98	  &250     &0.040         &0.5      &12      &0.03  \\
12  &W51 North	     	&160	  &90	   &0.068         &1.5      &13      &0.09  \\
13  &G24.78+0.08 A2  	&163	  &80	   &0.020         &0.75     &12      &0.03  \\
14  &W51e2	     	&241	  &140     &0.010         &2.0      &14, 15  &0.04  \\
15  &G24.78+0.08 A1  	&264	  &130     &0.020         &1.5      &12      &0.04  \\
16  &G23.01$-$0.41   	&274	  &380     &0.060$^k$     &0.6      &11      &0.03  \\
17  &IRAS~18566+0408     	&304	  &70	   &0.034         &3.0      &16      &0.09  \\
18  &G19.61$-$0.23   	&415	  &415     &0.031         &1.0      &6       &0.03  \\ 
19  &G31.41+0.31     	&508	  &490     &0.040         &2.1      &12      &0.06  \\
20  &NGC 7538S	     	&607	  &100     &0.070         &1.35     &17      &0.04  \\
\hline

\end{tabular}

 $^a$ Number in Fig.~\ref{minfall}. \\
 $^b$ Masses estimated assuming the dust opacities of Ossenkopf \& Henning~(\cite{ossen94}; see
 Sect.~\ref{toroids}). \\ 
 $^c$ Masses from the literature (see Sect.~\ref{toroids}). \\
 $^d$ References for the core parameters: 1: Cesaroni et al.~(\cite{cesa07}); 2: Cesaroni et
 al.~(\cite{cesa05}) 3: Patel et al.~(\cite{patel05}); 4: Beuther et al.~(\cite{beuther07c}); 5:
 Bernard et al.~(\cite{bernard99}); 6: This work; 7: Galv\'an-Madrid et al.~(\cite{galvan09}); 8:
 Hunter et al.~(\cite{hunter06}); 9: Beuther et al.~(\cite{beuther08});  10: Beuther et
 al.~(\cite{beuther05}); 11: Furuya et al.~(\cite{furuya08}); 12: Beltr\'an et
 al.~(\cite{beltran04}); 13: Zapata et al.~(\cite{zapata08}); 14: Zhang \& Ho~(\cite{zhang97}); 15:
 Shi et al.~(\cite{shi10}); 16: Zhang et al.~(\cite{zhang07});  17: Sandell et
 al.~(\cite{sandell03})\\
 $^e$ The free-fall times $t_{\rm ff}$ have been estimated from the masses obtained assuming the dust
 opacities of Ossenkopf \& Henning~(\cite{ossen94}). The uncertainty in $t_{\rm ff}/t_{\rm rot}$
 is $\sim$25\%.\\
 $^f$ Mass estimated using $\beta$=2 (Patel et al.~\cite{patel05}). \\
 $^g$ The mass of this core has been estimated from CS (Bernard et al.~\cite{bernard99}) and is
 the value used to derive $t_{\rm ff}$. \\ 
 $^h$ Mass estimated using $\beta$=1.5  (Galv\'an-Madrid et al.~\cite{galvan09}). \\
 $^i$ $V_{\rm rot}$ is $\sim$3--4~\kms\ (Galv\'an-Madrid et al.~\cite{galvan09}). \\
 $^j$ Mass estimated assuming a hot core temperature of 100~K (Beuther et al.~\cite{beuther05}).  \\
 $^k$ $R$ ranges from 0.055 to 0.068~pc depending on the observational wavelength (Furuya et
 al.~\cite{furuya08}). \\

\end{table*}

\section{Conclusions}

We have analyzed millimeter high-angular resolution data, obtained with the IRAM PdBI interferometer,
of the dust and gas emission towards the HMCs G10.62$-$0.38, G19.61$-$0.23, and G29.96$-0.02$. The
aims were to study the structure of the cores, map the molecular outflows powered by the YSOs embedded
in the HMCs, and reveal possible velocity gradients indicative of rotation. 

The continuum emission at 2.7~mm is clearly dominated by free-free emission from the \UC\
region(s), while at 1.4~mm dust emission from the HMCs prevails.

The CH$_3$CN~(12--11) LSR velocity maps reveal the existence of clear velocity gradients in the three
HMCs oriented perpendicular to the direction of the corresponding bipolar outflows.  The gradients
are interpreted as rotation. The gas temperatures, used to derive the mass of the
cores, have been obtained by means of the rotational  diagram method, and are in the range of
87--244~K. The diameters and masses of the toroids lie in the range of 4550--12600~AU, and
28--415~$M_\odot$, respectively. 

The masses of the cores are comparable to the corresponding virial masses suggesting that
turbulence could support the toroids. Given that the dynamical mass is considerably smaller than
the mass of the cores, we suggest that the toroids are not centrifugally supported and are
possibly undergoing collapse. For G19 and G29, this is also suggested by the redshifted
absorption seen in $^{13}$CO~(2--1). We infer that infall onto the embedded (proto)stars
proceeds with rates of $\sim$10$^{-2}$~$M_\odot$ yr$^{-1}$, and on timescales of the order of
$\sim$4$\times$10$^3$--10$^4$ yr.  The infall rates derived for G19 and G29 are two orders of
magnitude greater than the accretion rates indirectly estimated from the mass loss rate of the
corresponding outflows. This suggests that the material in the toroids is not infalling onto a
single massive star, responsible for the corresponding molecular outflow, but onto a cluster of
stars.


The masses, sizes, and infall rates of the rotating toroids are orders of magnitude higher than
those of real accretion disks around lower mass stars, while their lifetimes are much shorter. Also,
we find  that the higher the mass of the rotating structure, the lower the $t_{\rm ff}/t_{\rm rot}$
ratio is. Therefore,  circumstellar disks around B-type stars are centrifugally supported and undergo
Keplerian rotation, while massive toroids around O-type stars appear to be transient structures
undergoing collapse.

\begin{acknowledgements} 
We thank the staff of IRAM for their help during the
observations and data reduction. 
\end{acknowledgements}

\end{document}